\documentstyle[preprint,aps,pra]{revtex}
\tightenlines
\begin{document}

\draft
\def\pc{\protect\cite}
\def\hpsi{\hat\psi}
\def\tpsi{\tilde\psi}
\def\br{{\bf r}}
\def\bk{{\bf k}}
\def\bu{{\bf u}}
\def\bw{{\bf w}}
\def\brt{\br,t}
\def\bbrt{(\brt)}
\def\cphio{\Phi_0}
\def\beq{\begin{equation}}
\def\eeq{\end{equation}}
\def\bea{\begin{eqnarray}}
\def\eea{\end{eqnarray}}
\def\bna{\bbox{\nabla}}
\def\bp{{\bf p}}
\def\bv{{\bf v}}
\def\tn{\tilde n}
\def\tp{\tilde p}
\def\be{\bbox{\eta}}
\def\epst{\epsilon,t}
\def\intdrdp{\int {d{\bf r} d{\bf p} \over (2\pi\hbar)^3}}

\title{Condensate growth in trapped Bose gases}

\author{M. J. Bijlsma, E. Zaremba\footnote{Permanent address: 
Department of Physics, Queen's 
University, Kingston, Ontario, Canada K7L 3N6.},
and H. T. C. Stoof}
\address{Institute for Theoretical Physics, University of Utrecht,
\break Princetonplein 5, 3584 CC Utrecht, The Netherlands}

\maketitle

\begin{abstract}
We study the dynamics of condensate formation  
in an inhomogeneous 
trapped Bose gas with a positive interatomic scattering length.
We take into account both the nonequilibrium kinetics 
of the thermal cloud and 
the Hartree-Fock mean-field effects in the condensed
and the noncondensed parts of the gas.
Our growth equations are solved
numerically by assuming that the thermal
component behaves ergodically and that the condensate, treated 
within the Thomas-Fermi approximation, grows adiabatically.
Our simulations are in good qualitative agreement with experiment,
however important discrepancies concerning details of the growth
behaviour remain.
\end{abstract}


\newpage
\section{INTRODUCTION}
\label{section1}
The discovery of Bose-Einstein condensation (BEC) in trapped atomic 
vapors of $^{87}$Rb~\cite{anderson95}, $^{7}$Li~\cite{bradley95}, and
$^{23}$Na~\cite{davis95} has initiated a period of intense experimental
and theoretical activity. A great deal of information is now available
about the equilibrium properties of these novel 
systems~\cite{dalfovo99}, but much remains to be understood about their
nonequilibrium behavior. One of the most basic aspects
concerns the nonequilibrium growth of the condensate 
which occurs in the process of cooling a nondegenerate trapped Bose gas
to a final temperature below the BEC transition. This important problem
was addressed even before the first observation of BEC in trapped
atomic gases~\cite{levich77,snoke,stoof91}, 
and has interesting implications for the
general problem of second-order phase transitions, from superfluidity in
liquid $^{4}$He~\cite{hendry} to problems in cosmology~\cite{kibble}.

Up to now, the most detailed study of condensate formation 
was carried out using
a gas of $^{23}$Na atoms confined within a highly anisotropic 
cigar-shaped trap~\cite{stamper98}. In these experiments, 
the sodium atoms were evaporatively cooled
to a temperature just above the critical temperature
and subsequently quenched by applying a rapid rf sweep. 
The latter step removes all, or at least a large fraction, of
the atoms above a certain energy, after which the Bose gas relaxes 
to a new equilibrium state below the critical temperature.
The growth of the condensate during the equilibration process was
monitored using a nondestructive imaging technique which provided a
direct measure of the size of the condensate as a function of time. 
In this way, the
characteristic time scale for the growth of the condensate could be
determined, and for the particular system studied, was found to be
of the order of 100 ms.

A theoretical description of these experiments requires a theory that 
can account for the coupled nonequilibrium dynamics of both the 
noncondensed and condensed
components of a trapped Bose gas, and includes in particular
the collisional processes which transfer atoms between the two 
components. Thus far several such theories have been developed,
which roughly speaking fall into two categories.
One class of theories focuses on describing the 
dynamics of the average value of the order parameter for BEC, i.e.,
the condensate wave function, 
whereas the other incorporates also the fluctuations around 
this mean value.
The latter of course, becomes important when the fluctuations are large
compared to the mean value, i.e., close to the critical temperature. 
This is analogous to the situation in laser theory~\cite{mandel}.

A theory that describes both the average value for
the order parameter as well its fluctuations
can be obtained in two, essentially 
equivalent ways. First, one can start from a master equation
for the many-body density matrix and derive an equation 
of motion for the one-particle density matrix
by means of a perturbative treatment of the interactions. This was 
the route followed by Gardiner and Zoller~\cite{gardiner97II},
in a series of papers. 
Second, one can use field-theoretic methods to obtain a 
nonperturbative Fokker-Planck equation that describes the 
nonequilibrium dynamics of the gas. This was the formulation 
developed by Stoof~\cite{stoof99}. 
These two approaches in principle yield
a description of the nonequilibrium dynamics
that is capable of obtaining the complete 
probability distribution for the  order parameter. 

Alternatively, a theory describing the dynamics of the
mean-field value for the macroscopic wave function 
can also be obtained in  a straightforward decoupling approach,
which has been implemented 
by Kirkpatrick and Dorfmann~\cite{kirkpatrick85},
Proukakis {\it et al.}~\cite{proukakis},
Walser {\it et al.}~\cite{walser99}, and in most detail 
for the trapped case by Zaremba, Nikuni, and Griffin~\cite{zaremba99}.
In this approach, one assumes the order parameter
to be nonzero at all temperatures,
and decouples the hierarchy of equations of motion
that exists for the correlation functions of the 
second-quantized field operators. Thus one again obtains
a perturbative expansion for these equations of motion.

The first quantitative calculations of condensate growth for
trapped Bose gases were carried
out by Gardiner {\it et al.}~\cite{gardiner97}, 
and although good qualitative agreement with experiment was found, a
number of quantitative discrepancies remained. 
For example, the reported experimental growth rates were 
up to a factor 30 larger than the initial theoretical results, 
and had a temperature dependence opposite to that 
predicted~\cite{stamper98}. By removing some simplifying
approximations in subsequent calculations, the theoretical results 
were improved, but discrepancies of up to a factor of 3 still 
remained in some cases.
From a purely theoretical point of view, one can attribute some of these
discrepancies to the approximations made in the calculations.
First, the dynamics of the noncondensate was 
to a large extent neglected. 
Although the time evolution of the occupancy of low-lying
states was included in the simulations, the high energy states were
represented by an equilibrium particle reservoir having a fixed chemical
potential. This latter assumption is inconsistent
with the nonequilibrium initial state established by the experimental 
quench procedure. Second, the effect of the mean field of the condensate
on the noncondensate was included rather crudely by a linear
rescaling of the low-energy density of states of the noncondensed 
atoms.

Our aim in the present paper is to improve on these calculations by
taking fully into account the relaxational dynamics of the thermal, or
noncondensed,
component that takes place in the presence of the mean field of the
condensate. We do this by starting from the above mentioned theories
describing the growth process, which
provide us with a nonlinear Schr\"odinger equation for
the condensate and a kinetic equation for the thermal component.
This coupled set of equations is still difficult to deal with
and a number of physically motivated approximations are made to simplify
the problem. We assume that the condensate grows adiabatically, having 
an equilibrium spatial distribution determined by the instantaneous
number of atoms in the condensate. This assumption is also made in
earlier work~\cite{gardiner97}. The noncondensate is treated by solving
a semiclassical Boltzmann 
equation~\cite{stoof99,kirkpatrick85,zaremba99,eckern}
in the ergodic approximation, which again has been used previously by 
numerous authors~\cite{snoke,luiten96,holland97,jaksch}.
These assumptions allow us to obtain numerically a detailed description
of the growth of the condensate, including the effects of both the 
dynamics of the thermal cloud and
its mean-field interaction with the condensate.

The paper is organized as follows. 
In Sec.~\ref{section2} we summarize the
theory of the nonequilibrium dynamics
of a trapped Bose gas as developed previously. In Sec.~\ref{section3}
we introduce the central assumption, the ergodic approximation, that 
allows us to numerically solve the Boltzmann equation.
In addition, we briefly discuss
the adiabatic approximation for the condensate. 
In Sec.~\ref{section4} we treat in some detail 
particle number and energy conservation.
Sec.~\ref{section5} introduces the Thomas-Fermi 
approximation and gives some analytical results for the density of
states and other quantities of interest. The numerical solution of our
kinetic equations is discussed in Sec.~\ref{section6} and our results
for the growth of a condensate are presented.
We end in Sec.~\ref{section7} with a discussion and an outlook.  

\section{NONEQUILIBRIUM DYNAMICS}
\label{section2}
As discussed in the previous section, the nonequilibrium dynamics of a
trapped Bose gas is governed by a set of equations for the condensate
and noncondensate components. These equations have been presented in 
various 
forms~\cite{gardiner97,stoof99,kirkpatrick85,proukakis,walser99,zaremba99}, 
but they all
describe the coherent dynamics of the gas due to mean-field
interactions, as well as the incoherent dynamics associated with atomic
collisions. The equations and notation we use are taken from
Refs.~\onlinecite{stoof99,zaremba99}.

The noncondensate is treated using a semiclassical Boltzmann equation 
for the phase space distribution function $f({\bf r},{\bf p},t)$. 
This semiclassical description is justified when the largest 
level spacing in  the external trapping potential is small compared to 
the thermal excitation energy. Moreover, mean-field interactions are
included at the level of the Hartree-Fock approximation. In this 
situation, the quantum kinetic equation for the thermal excitations
takes the form~\cite{stoof99,zaremba99}

\bea
\label{eq1}
{\partial f({\bf r},{\bf p},t) \over \partial t} + {{\bf p} \over m} 
\cdot \bbox{\nabla} f({\bf r},{\bf p},t)- \bbox{\nabla} U({\bf r},t) \cdot
\bbox{\nabla}_{{\bf p}} f({\bf r} ,{\bf p},t)  = 
C_{12}[f] + C_{22}[f] \; .
\eea
Here, the effective potential $U({\bf r},t)\equiv 
U_{\rm ext}({\bf r})+2g [n_c({\bf r},t)+\tilde n({\bf r},t)]$
is the sum of the external trapping potential $U_{\rm ext}$ and
the self-consistent Hartree-Fock mean field. The latter is determined by
the condensate density $n_c({\bf r},t)$, defined below, and 
the noncondensate density $\tilde n({\bf r},t)$ given by

\bea
\label{eq2}
\tilde n({\bf r},t)=\int\frac{d{\bf p}}{(2 \pi \hbar)^3}
f({\bf r},{\bf p},t) \; .
\eea
As usual, we treat the interactions in the $s$-wave approximation 
which results in the bare interaction being replaced by a  
contact interaction with an effective coupling constant 
$g=4\pi \hbar^{2} a/m$ proportional to the $s$-wave scattering
length $a$. The effective coupling constant is in fact equal to the 
two-body $T$-matrix, and to emphasize this connection, it is 
denoted by $T^{2B}$ in some works~\cite{stoof99}.
The collision terms appearing in Eq.~(\ref{eq1}) are given by

\bea
\label{eq3}
C_{22}[f] & \equiv & {4 \pi \over \hbar} g^2
\int{d{\bf p}_2\over(2\pi\hbar)^3}
\int{d{\bf p}_3\over(2\pi\hbar)^3}
\int{d{\bf p}_4\over(2\pi\hbar)^3}
(2\pi\hbar)^3\delta ({\bf p}+{\bf p}_2 
-{\bf p}_3 -{\bf p}_4)\nonumber\\
& & \times \delta(E+E_{2}-E_{3}-E_{4})
\left[(1+f)(1+f_2)f_3f_4-ff_2(1+f_3)(1+f_4)\right] \; ,
\end{eqnarray}
\begin{eqnarray}
\label{eq4}
C_{12}[f]& \equiv & {4 \pi \over \hbar} g^2
n_c 
\int\frac{d{\bf p}_2}{(2\pi\hbar)^3}
\int\frac{d{\bf p}_3}{(2\pi\hbar)^3} 
\int\frac{d{\bf p}_4}{(2\pi\hbar)^3}
(2\pi\hbar)^3\delta(m{\bf v}_c+{\bf p}_2-{\bf p}_3-{\bf p}_4) \cr
& & \times
\delta(E_c+E_{2}-E_{3}-E_{4})
(2\pi\hbar)^3
[\delta({\bf p}-{\bf p}_2)-\delta({\bf p}-{\bf p}_3)
-\delta({\bf p}-{\bf p}_4)] \cr
& &\times [(1+f_2)f_3f_4-f_2(1+f_3)(1+f_4)] \; ,
\eea
with $f \equiv f({\bf r, p}, t)$, and $f_i\equiv f({\bf r, p}_i, t)$.
We note that Eq.~(\ref{eq4}) takes into account the fact that a
condensate atom locally has an energy

\bea
E_{c}({\bf r},t)=\mu_{c}({\bf r},t)+
{1 \over 2} m {\bf v}_{c}^{2}({\bf r},t) \; ,
\eea
a momentum $m {\bf v}_c({\bf r},t)$, 
and a chemical potential $\mu_c({\bf r},t)$.
These quantities are defined explicitly below.
In addition, the energy of a noncondensate atom
in the Hartee-Fock approximation is
\bea
E(\br,\bp,t)=\frac{{\bf p}^2}{2m}+U({\bf r},t) \; .
\eea
The energy variables $E_i$ appearing in Eqs.~(\ref{eq3}) and 
(\ref{eq4}) are defined as $E_{i}=E(\br,\bp_{i},t)$.

In contrast to the thermal cloud, the dynamics of the condensate is
determined by a time-dependent dissipative nonlinear Schr\"odinger 
equation~\cite{stoof99,zaremba99},

\bea
\label{nlse}
i \hbar \frac{\partial \Psi(\br,t)}{\partial t} & = &
\left\{
-\frac{\hbar^{2} \bbox{\nabla}^{2}}{2m} + U_{\rm ext}(\br) +
g\left[ 2 \tilde n(\br,t) + n_{c}(\br,t) \right]
\frac{}{\mbox{}} \!\!  
- i R(\br,t)  
\right\} \Psi(\br,t) \; ,
\eea
where the dissipative term, i.e., $R({\bf r},t)$,  is given by

\bea
R & \equiv & 
{\hbar\over 2n_c}\int {d {\bf p}\over (2\pi\hbar)^3}C_{12}[f]\cr
&=& {g^2 \over (2\pi)^5\hbar^6} 
\int \prod_{i=1,4}d{\bf p}_i\,
\delta(m{\bf v}_c+{\bf p}_2-{\bf p}_3-{\bf p}_4) \cr
& & \qquad\times
\delta(E_c+E_2 -E_3-E_4)
[\delta({\bf p}_1-{\bf p}_2)-\delta({\bf p}_1-{\bf p}_3)
-\delta({\bf p}_1-{\bf p}_4)] \cr
& &\qquad\times [(1+f_2)f_3f_4-f_2(1+f_3)(1+f_4)] \,.
\eea
The appearance of this dissipative term in Eq.~(\ref{nlse}) is a
consequence of the collisional processes, described by $C_{12}$, which 
have the effect of transferring particles between the condensate and 
noncondensate.
The dissipative term  is needed in order to ensure
overall particle number conservation of the entire system.
At a more fundamental level, the condensate wave function is determined
by taking the expectation value of a Bose field with respect 
to a probability distribution that satisfies the Fokker-Planck equation
mentioned previously~\cite{stoof99}. The Langevin equation one 
derives within this formulation has the form of a dissipative nonlinear 
Schr\"odinger equation with a noise term. It only reduces to
Eq.~(\ref{nlse}) in a mean-field approximation. This points to the need
for exercising care in interpreting the order parameter
occuring in the nonlinear Schr\"odinger equation as
the condensate wave function. Due to the underlying $U(1)$ gauge
invariance associated with strict particle number conservation,
the expectation value is in fact always equal to zero as a result of 
the diffusion of the global phase of the condensate wave function. 
In first instance this effect can be neglected and we are then
effectively treating the system as if the $U(1)$ gauge invariance is 
explicitly broken.

It is convenient to rewrite  Eq.~(\ref{nlse}) in terms
of amplitude and phase variables defined by 
$\Psi(\br,t)=\sqrt{n_{c}(\br,t)} \exp[i \theta(\br,t)]$. Substituting 
this form of the wave function into Eq.~(\ref{nlse}), we obtain

\bea
\label{eq9}
{\partial n_{c}(\br,t) \over \partial t} + 
\bna \left[
{\bf v}_{c}(\br,t) n_{c}(\br,t)
\right] = - {2 \over \hbar} R(\br,t) n_{c}(\br,t) \; ,
\eea
and

\bea
m {\partial {\bf v}_{c}(\br,t) \over \partial t} +
\bna \left[ \mu_{c}({\br,t}) + 
{m {\bf v}_{c}(\br,t)^{2} \over 2} \right] = 0 \; .
\eea
Here, we have defined the local chemical potential and superfluid
velocity by

\bea
\mu_{c}(\br,t) =
U_{\rm ext}(\br) + g \left[
n_{c}(\br,t) 
+ 2 \tilde n(\br,t) \right]
- {\hbar^{2} \over 2 m} 
{\bna^{2} \sqrt{n_{c}(\br,t)} \over \sqrt{n_{c}(\br,t)}} \; ,
\eea
and

\bea
{\bf v}_{c}(\br,t) = {\hbar \over m} \bna \theta(\br,t) \; ,
\eea
respectively. It can easily be shown that this set of equations for
the condensate and thermal cloud is consistent with the conservation of
the total number of particles in the system~\cite{zaremba99}.

\section{ERGODIC APPROXIMATION}
\label{section3}
Our main objective in this paper is to apply the kinetic theory 
formulated above to the
problem of condensate formation. In order to make progress we introduce 
a number of additional approximations. The first and most essential,
is the assumption of ergodicity~\cite{snoke,luiten96,holland97,jaksch} 
which has been widely used in the literature on kinetic theory. 
This assumes that equilibration of atoms within one energy level 
occurs on a much shorter time scale than equilibration of atoms 
between different energy levels. With this assumption, 
all points in phase space having the same energy are equally probable,
and the distribution function therefore only depends on the phase space
variables through the energy variable $E(\br,\bp,t)$, i.e., 
$f(\br,\bp,t) \equiv g(E(\br,\bp,t),t)$. 
In equilibrium this is certainly correct, but
the assumption requires justification 
for any particular nonequilibrium application. 
Unfortunately, we are not aware of any explicit checks that have been
made which might indicate that the assumption is correct 
for the situations we wish to consider. Nevertheless, it appears to
be physically reasonable that
for quantities that vary on a time scale of the order of several
collision times, the approximation is sufficiently accurate.

The ergodic approximation allows us to derive a simplified kinetic
equation for the energy distribution function $g(\epst)$. 
This is accomplished by means of the relation 

\beq
\rho(\epst)g(\epst) \equiv \intdrdp
\delta(\epsilon-E(\br,\bp,t))f(\br,\bp,t) \, ,
\eeq
which shows that the phase-space projection defined on the right-hand 
side yields the product of $g(\epst)$ and the density of states

\bea
\label{eq14}
\rho(\epst) &=& \intdrdp \delta(\epsilon-E(\br,\bp,t))\nonumber\\
&=& {m^{3/2} \over \sqrt{2} \pi^2 \hbar^3} \int_{U\le \epsilon} 
d \br 
\sqrt{\epsilon-U(\br,t)}\,.
\eea
We note that the density of states is defined on the variable energy 
range $U_{\rm min}(t) \le \epsilon < \infty$ where $U_{\rm min}(t)$ 
is the minimum value of $U(\brt)$ at time $t$. The time dependence of 
the density of states is one of the aspects distinguishing the
present development from previous work~\cite{luiten96,holland97}.

We now apply the phase-space projection to the kinetic equation in 
Eq.~(\ref{eq1}). As a result of this operation,
the streaming terms in the Boltzmann equation, i.e., the second and 
third terms on the left-hand side of Eq.~(\ref{eq1}), 
cancel each other. Only the projection of the time-derivative term 
survives. This results in

\beq
\label{eq15}
\intdrdp \delta(\epsilon-E(\br,\bp,t)) {\partial f(\br,\bp,t) \over 
\partial t} = \rho(\epst){\partial g(\epst)\over \partial t} 
+ \rho_{\rm w}(\epsilon,t)
{\partial g(\epst)\over \partial \epsilon} \, .
\eeq
Here, we have introduced a {\it weighted} density of states

\bea
\label{eq16}
\rho_{\rm w}(\epsilon,t) & = & \intdrdp \delta(\epsilon-E(\br,\bp,t))
{\partial U(\brt)\over \partial t} \nonumber \\
& = & {m^{3/2} \over \sqrt{2} \pi^2 \hbar^3} \int_{U\le \epsilon} 
d \br \sqrt{\epsilon-U(\br,t)}\,
{\partial U(\brt)\over \partial t} \,.
\eea
This quantity depends explicitly on the time derivative of
the noncondensate potential which in turn is determined by the time
derivatives of both the
condensate and noncondensate densities. Some formal details regarding
its evaluation are given in the Appendix.

Noting that 

\beq
{\partial \rho(\epst) \over \partial t} = -{\partial \rho_{\rm w}(\epst) 
\over \partial \epsilon}\,,
\eeq
Eq. (\ref{eq15}) can be written as

\beq
\label{eq18}
\intdrdp\, \delta(\epsilon-E(\br,\bp,t)) {\partial f(\br,\bp,t) \over 
\partial t} = {\partial \over \partial t}\left ( \rho g \right )  
+ {\partial \over \partial \epsilon}\left ( \rho_{\rm w} g \right )\,.
\eeq 
We thus arrive at the projected kinetic equation

\beq
\label{eq19}
{\partial \over \partial t}\left ( \rho g \right )  
+ {\partial \over \partial \epsilon}\left ( \rho_{\rm w} g \right )
=I_{12}+I_{22}\,,
\eeq 
where the phase space projections of the collision integrals are 
defined as

\begin{mathletters}
\label{eq20}
\bea
I_{12}(\epsilon,t) &\equiv&\intdrdp
\delta(\epsilon - E(\br,\bp,t)) C_{12}[f]
\eea
\bea
I_{22}(\epsilon,t) &\equiv&\intdrdp
\delta(\epsilon - E(\br,\bp,t)) C_{22}[f] \, .
\eea
\end{mathletters}
The result in Eq.~(\ref{eq19}) is the kinetic equation 
that we solve numerically.

We now derive in some detail explicit
expressions for the collision integrals in Eq.~(\ref{eq20}).
Although an expression for $I_{22}$ was given in earlier 
work~\cite{luiten96}, we present here an alternative derivation
which can also be adapted to the case of the $I_{12}$ collision 
integral. For the $I_{22}$ collision integral we have

\bea
\label{eq21}
I_{22}(\epsilon_1,t)
&=& {4\pi g^2 \over (2\pi)^9\hbar^{10}}\int 
d\epsilon_2 \int d\epsilon_3
\int d\epsilon_4\,\delta(\epsilon_1+\epsilon_2-\epsilon_3-\epsilon_4)
\nonumber \\ &&\times
\left [ (1+g_1)(1+g_2)g_3 g_4 - g_1 g_2 (1+g_3)(1+g_4)
\right ]\nonumber \\
&&\times \int d{\bf r} \left( \prod_{i=1,4} \int d\bp_{i} \right) \,
\delta(\bp_1+\bp_2-\bp_3-\bp_4)\nonumber \\
&&\qquad\qquad\qquad \times \delta(\epsilon_1-E_1)
\delta(\epsilon_2-E_2)\delta(\epsilon_3-E_3)\delta(\epsilon_4-E_4) \; ,
\eea
where we have introduced the short-hand 
notation $g_i = g(\epsilon_i,t)$.
We consider first the momentum integrals in Eq. (\ref{eq21}) which,
with the replacement $\bp_3 \to -\bp_3$ and $\bp_4 \to -\bp_4$, can be
written as

\bea
\label{eq22}
J_{22} &\equiv& \int d{\bf p}_1\int d{\bf p}_2\int d{\bf p}_3
\int d{\bf p}_4\, \delta(\bp_1+\bp_2+\bp_3+\bp_4)\nonumber \\
&&\times \delta(\epsilon_1-E_1)
\delta(\epsilon_2-E_2)\delta(\epsilon_3-E_3)\delta(\epsilon_4-E_4)
\nonumber \\
&=& \int {d\bbox{\xi} \over (2\pi)^3} \prod_{i=1,4} \int d\bp_i
\, e^{i\bbox{\xi}\cdot\bp_i}\delta(\epsilon_i-E_i)\,.
\eea
In obtaining this expression, we have introduced a Fourier
representation of the momentum conserving delta function. Performing
the integrals in Eq. (\ref{eq22}) 
with respect to the momentum variables,
we obtain

\beq
J_{22}=(4\pi m)^4 \left (\prod_{i=1}^4 \theta(\epsilon_i
-U) \right ) \int {d\bbox{\xi} \over (2\pi)^3}
{\sin{\xi p_1}\sin{\xi p_2}\sin{\xi p_3}\sin{\xi p_4}\over \xi^4} \; ,
\eeq
where now it is understood that $p_i=\sqrt{2m(\epsilon_i-U)}$.
The product of theta functions can be replace by $\theta(\epsilon_{\rm
min}-U)$, with $\epsilon_{\rm min}$ the minimum value of the four
energy variables.
Performing the remaining integral with respect to the $\bbox{\xi}$
variable, we find

\bea
\label{eq24}
J_{22}=(2\pi)^3 m^4 \theta(\epsilon_{\rm min} -U)
&\Big [& |p_1-p_2+p_3+p_4|-|p_1-p_2+p_3-p_4|\nonumber \\
&+&|p_1-p_2-p_3-p_4|- |p_1-p_2-p_3+p_4|\nonumber \\
&+&|p_1+p_2+p_3-p_4|-|p_1+p_2-p_3-p_4|\nonumber \\
&+& |p_1+p_2-p_3+p_4| -|p_1+p_2+p_3+p_4|\, \Big ]\,.
\eea
This expression is valid for arbitrary values of the momenta but
simplifies when energy conservation is taken into account.
Since the energy conserving delta function 
$\delta(\epsilon_1+\epsilon_2-\epsilon_3-\epsilon_4)$ in
Eq.~(\ref{eq21}) imposes the constraint $p_1^2+p_2^2=p_3^2+p_4^2$, 
Eq.~(\ref{eq24}) can be reduced to 

\beq
J_{22}=4(2\pi)^3 m^4 \theta(\epsilon_{\rm min} -U) 
\sqrt{2m(\epsilon_{\rm min}-U)} \,.
\eeq
Substituting this expression for $J_{22}$ into
Eq. (\ref{eq21}), we finally obtain  

\bea
\label{I22}
I_{22}(\epsilon_1,t) &=& 
{m^3 g^2 \over 2\pi^3\hbar^7}\int d\epsilon_2 \int d\epsilon_3
\int d\epsilon_4 \,\rho(\epsilon_{\rm min}) 
\delta(\epsilon_1+\epsilon_2-\epsilon_3-\epsilon_4) 
\nonumber \\ &&\times
\left [ (1+g_1)(1+g_2)g_3 g_4 - g_1 g_2 (1+g_3)(1+g_4)
\right ] \; ,
\eea
where we have used the definition of the density of states in 
Eq.~(\ref{eq14}). This is precisely the result obtained by Snoke
and Wolfe~\cite{snoke} and  
Luiten, Reynolds, and Walraven~\cite{luiten96}, using a different method.
We note that if all energies are expressed in units of $\hbar \bar
\omega$, the $I_{22}$ integral has an overall factor of $(a/l)^2\bar
\omega$ where $l=\sqrt{\hbar/m\bar\omega}$ is the average harmonic
oscillator length. This factor defines a characteristic time which
can be used as the time unit in the simulations.

The $I_{12}$ collision integral can be dealt with in a similar way 
if the superfluid velocity ${\bf v}_{c}$ in Eq.~(\ref{eq4}) is set
to zero. The validity of this approximation follows from our assumption
that the condensate grows adiabatically. The 
magnitude of the superfluid 
velocity ${\bf v}_{c}$ is then typically of the order of 
$\dot R(t)$, where $R(t)$ is the
radius of the condensate. This velocity is
small compared to the characteristic velocities ${\bf p}/m \approx 
\sqrt{2 k_B T / m}$ of the thermal atoms participating in a collision,
which justifies the neglect of $m\bv_c$ in Eq.~(\ref{eq4}).
The expression for $I_{12}$ then reads

\bea
\label{eq28}
I_{12}(\epsilon_1,t) &=& {4 \pi g^2 \over (2\pi)^6 \hbar^7} 
\int d \br \, n_c(\brt) \int d\epsilon_2 \int d\epsilon_3
\int d\epsilon_4\, \delta(E_c({\bf r},t)+\epsilon_2-\epsilon_3-\epsilon_4)
\nonumber \\ &&\times
\left [ \delta(\epsilon_1-\epsilon_2)-\delta(\epsilon_1-\epsilon_3)
-\delta(\epsilon_1-\epsilon_4) \right ]
\left [ (1+g_2)g_3 g_4 - g_2 (1+g_3)(1+g_4)
\right ]\nonumber \\
&&\times \int d\bp_2 \int d\bp_3 \int d\bp_4\,
\delta(\bp_2-\bp_3-\bp_4)
\delta(\epsilon_2-E_2)\delta(\epsilon_3-E_3)\delta(\epsilon_4-E_4) \; .
\eea
If we now define $J_{12}$ analogously to $J_{22}$, we have
 
\bea
J_{12} &\equiv& \int d\bp_2 \int d\bp_3 \int d\bp_4 \,
\delta(\bp_2+\bp_3+\bp_4)
\delta(\epsilon_2-E_2)\delta(\epsilon_3-E_3)\delta(\epsilon_4-E_4)
\nonumber \\
&=& \int {d\bbox{\xi} \over (2\pi)^3} \prod_{i=2}^4 \int d\bp_i \,
e^{i\bbox{\xi} \cdot\bp_i} \delta(\epsilon_i-E_i)\nonumber \\
&=& (4\pi m)^3 \theta(\epsilon_{\rm min}-U)
\int {d\bbox{\xi} \over (2\pi)^3} {\sin \xi p_2 \sin
\xi p_3 \sin \xi p_4
\over \xi^3}\nonumber \\
&=&8 \pi^2 m^3 \theta(\epsilon_{\rm min}-U) S(p_2,p_3,p_4) \, ,
\eea
where $\epsilon_{\rm min}$ is the minimum value of
$\epsilon_{2}$, $\epsilon_{3}$, and $\epsilon_{4}$, and 

\bea
\label{Sp2p3p4}
S(p_2,p_3,p_4) \equiv {1 \over 2} \left[ {\rm sgn}(p_2+p_3-p_4) 
\right. & + & {\rm sgn}(p_2-p_3+p_4) \nonumber \\
- {\rm sgn}(p_2+p_3+p_4) & - & \left. 
{\rm sgn}(p_2-p_3-p_4) \right] \; .
\eea
Note that this is a boolean function which 
takes on values of 0 and 1.
Inserting the expression for $J_{12}$ into Eq.~(\ref{eq28}), 
we finally obtain for the $I_{12}$ collision integral the result

\bea
\label{I12}
I_{12}(\epsilon_1,t) &=&
{m^3 g^2 \over 2\pi^3\hbar^7} 
\int d\epsilon_2 \int d\epsilon_3
\int d\epsilon_4\, 
\left[ \delta(\epsilon_1-\epsilon_2)-\delta(\epsilon_1-\epsilon_3)
-\delta(\epsilon_1-\epsilon_4) \right]
\nonumber \\ &&\times
\left [ (1+g_2)g_3 g_4 - g_2 (1+g_3)(1+g_4)
\right ]\nonumber \\ &&\times 
\int_{U\le \epsilon_{\rm min}} d\br\, n_c(\brt) S(p_2,p_3,p_4)\,
\delta(E_c(\br,t)+\epsilon_2-\epsilon_3-\epsilon_4)\,.
\eea
A comparison of this expression with $I_{22}$ in Eq.~(\ref{I22}) shows
that the remaining spatial integral acts as an effective density of
states for scattering into the condensate. It can be evaluated
analytically in the Thomas-Fermi approximation for the condensate, as
shown in Sec.~\ref{section5}. The kinetic equation in Eq.~(\ref{eq19})
and the projected collision integrals in Eqs.~(\ref{I22}) and 
(\ref{I12}) are the main results of this section.

Before closing this section we point out one difficulty encountered 
when a numerical solution of Eq.~(\ref{eq19}) is attempted. As discussed
following Eq.~(\ref{eq14}), the time
dependence of the mean field potential $U(\br,t)$ implies that the
density of states in Eq.~(\ref{eq14}), and hence the energy distribution
function $g(\epsilon,t)$, are defined on a variable energy range. To
eliminate this variation, it is convenient to introduce the shifted
energy variable

\beq
\label{variable change}
\bar \epsilon \equiv \epsilon - U_{\rm min}(t) \, ,
\eeq
which leads to a fixed energy range
$0 \le \bar \epsilon < \infty$.
The density of states in terms of this new energy variable
is given by $\rho(\epsilon,t) = \rho(\bar\epsilon+U_{\rm min},t) \equiv
\bar\rho(\bar\epsilon,t)$. 
With $\bar\epsilon$ and $t$ as independent variables, the kinetic 
equation in Eq.~(\ref{eq19}), can be rewritten as

\beq
\label{eq44}
{\partial \over \partial t}\left ( \bar\rho \bar g \right )  
+ {\partial \over \partial \bar\epsilon}\left ( \bar\rho_{\rm w} \bar g 
\right ) =\bar I_{12}+\bar I_{22}\,.
\eeq 
Here, both $\bar\rho(\bar\epsilon,t)$ and
$\bar\rho_{\rm w}(\bar\epsilon,t)$ are 
defined by making the replacement
 
\beq
\label{eq43}
U(\brt) \to U(\brt) - U_{\rm min}(t) \equiv \overline U(\brt)\,, 
\eeq
in Eqs.~(\ref{eq14}) and (\ref{eq16}). Similarly, from
the definition of the collision integrals in Eqs.~(\ref{eq3}) 
and (\ref{eq4}), it can also be seen that the change of energy variable 
leads to the replacement of $U(\brt)$ by $\overline U(\brt)$
in this case as well. Thus, the
final kinetic equation in terms of the $\bar \epsilon$ variable
is unchanged in form from the original equation. We will henceforth drop
the overbar on the functions defined in terms of $\bar\epsilon$, with 
the understanding that the shifted potential $\overline U(\brt)$ is to 
be used wherever the potential appears in the original expressions. 
The possibility of using a fixed energy range in the solution of the
kinetic equation simplifies the numerical calculations considerably.

\section{COLLISIONAL INVARIANTS}
\label{section4}
In this section we explicitly consider two important 
quantities that should be conserved as the Bose gas 
condenses and equilibrates, namely, the total
number of particles and the total energy of the trapped
Bose gas. Together, they determine the final equilibrium state
of the  Bose-condensed gas, i.e., the number of particles
in the condensate, its chemical potential, and the 
temperature of the vapor.

\subsection{Particle Number Conservation}
The time rate of change of the total number of particles 
consists of the time rate of change of the number of condensed 
particles plus the time rate of change of the number of noncondensed 
particles.  Because the number of noncondensed particles is given by
$\tilde N(t) = (2 \pi \hbar)^{-3} 
\int d\br d\bp f(\bp,\br,t) = \int d\epsilon \rho(\epsilon)
g(\epsilon)$, the time rate of change of $\tilde N(t)$
can be found by 
integrating Eq.~(\ref{eq19}) over energy. We thus find
that 

\beq
\label{eq35}
{\partial \tilde N(t) \over \partial t} = \int d\epsilon \, 
I_{12}(\epsilon,t)\,, 
\eeq
where it is easily checked from Eq.~(\ref{I22}) that

\begin{eqnarray} 
\label{eq34}
\int d\epsilon \, I_{22}(\epsilon,t) = 0\,.
\end{eqnarray}
Note that we have assumed here that $\lim_{\epsilon \to U_{\rm min}} 
\rho_{\rm w}(\epsilon)g(\epsilon) = 0$. A finite limiting value can
arise if $g(\epsilon)$ approaches an equilibrium Bose distribution 
with a chemical potential $\mu = U_{\rm min}$ at long times, together 
with a weighted density of states which depends linearly on
$\epsilon - U_{\rm min}$ for energies close to $U_{\rm min}$. 
However, at any finite time in the growth
process, it is safe to use the zero limiting value. This is always the
case when the equilibrium chemical potential lies below 
$U_{\rm min}$.

To get the time rate of change of the total
number of condensate particles, we integrate
the continuity equation, Eq.~(\ref{eq9}), over space to find

\begin{eqnarray}
\label{eq36}
{\partial N_{c} \over \partial t} & = & 
- {2 \over \hbar} \int d{\bf r} R({\bf r},t) n_{c}({\bf r,t})
\nonumber \\
& = & - \int d\epsilon \, I_{12}(\epsilon,t)\,.
\end{eqnarray}
Combining this with Eq.~(\ref{eq35}) leads to

\begin{eqnarray}
{\partial (\tilde N + N_{c}) \over \partial t} & = & 0 \; ,
\end{eqnarray}
which demonstrates that the total number of particles is indeed 
conserved.

\subsection{Energy Conservation}
We now consider the conservation of the total energy of the system.
The total energy is given by

\bea
\label{eq38}
E_{\rm tot} & = & 
\intdrdp \left\{ {{\bf p}^2 \over 2m} +
U_{\rm ext}(\br) + g\left[
\tilde n(\br,t) + 2 n_{c}(\br,t) \right] \right\} f(\br,\bp,t) 
\nonumber \\
& & +
\int d\br \, \Psi^*(\brt) 
\left[ -{\hbar^2 \bbox{\nabla}^2\over 2m} + 
U_{\rm ext}(\br) + {g \over 2}
n_{c}(\br,t) \right] \Psi(\brt) \,.
\eea
The first term is the semi-classical expression for the total energy 
of the noncondensate. It contains the kinetic and external
potential energy, and the  Hartree-Fock
mean-field interaction energy of the noncondensed cloud interacting 
with itself and with the condensate. We note that the self-interaction
term is reduced by a factor of two relative to the condensate term to
avoid double counting this contribution.

The second term in Eq.~(\ref{eq38}) is the total energy of the 
condensate which contains the wave function $\Psi(\brt)$ with 
normalization

\beq
\int d\br \, |\Psi(\brt)|^2 = N_c(t) \, .
\eeq
It consists of the kinetic energy, the potential energy, and
the mean-field energy due to the interaction of the condensate
with itself. The mean-field interaction of the condensate
with the noncondensate
has already  been included in the expression for the energy 
of the noncondensed cloud. We now show
that this total energy is indeed conserved during the growth process.

Taking the time derivative of Eq.~(\ref{eq38})
leads to the following expression,

\bea
\label{eq40}
{\partial E_{\rm tot} \over \partial t} 
& = &
\intdrdp \left\{ {{\bf p}^2 \over 2m} +
U_{\rm ext}(\br) + 2 g \left[
\tilde n(\br,t) + n_{c}(\br,t) \right] \right\} 
{\partial f(\br,\bp,t) \over \partial t} 
\nonumber \\
& & + \int d\br \, {\partial \Psi^*(\brt) \over \partial t} 
\left\{ -{\hbar^2 \bbox{\nabla}^2\over 2m} + 
U_{\rm ext}(\br) + g
\left[
n_{c}(\br,t) + 2 \tilde n(\br,t) 
\right] \right\} \Psi(\brt) \nonumber \\
& & + \int d\br \, \Psi^*(\brt) 
\left\{ -{\hbar^2 \bbox{\nabla}^2\over 2m} + 
U_{\rm ext}(\br) + g
\left[
n_{c}(\br,t) + 2 \tilde n(\br,t) 
\right] \right\} {\partial \Psi(\brt) \over \partial t}  
\eea
The first term in Eq. (\ref{eq40}) can be rewritten as

\bea
\label{eq41}
\intdrdp E(\br,\bp,t) 
{\partial f(\br,\bp,t) \over \partial t} 
& = &
\intdrdp E(\br,\bp,t) \left( C_{12}[f] + C_{22}[f] \right)
\nonumber \\
& = & 
\int d\br E_{c}(\br,t) \int {d\bp \over (2\pi\hbar)^{3}}
C_{12}[f]\,,
\eea
where, to obtain this result, we have used the kinetic equation, 
Eq.~(\ref{eq1}), and the fact that the $C_{22}$ collision integral 
conserves energy.

If we  again assume that the condensate grows adiabatically 
as atoms are fed into it from the noncondensate, 
the condensate wavefunction $\Psi({\bf r},t)$
is a solution of the instantaneous Gross-Pitaevskii equation 

\bea
\label{eq42}
\left\{ - {\hbar^{2} \bbox{\nabla}^{2} \over 2 m} +
U_{\rm ext} + g \left[
n_{c}({\bf r}, t) + 2 \tilde n({\bf r}, t) \right]
\right\} \Psi(\br,t) = E_{c}(t) \Psi(\br,t) 
\, ,
\eea
with a time-dependent energy eigenvalue $E_{c}(t)$. For this spatially
independent condensate energy, Eq.~(\ref{eq41}) reduces to

\beq
\intdrdp E(\br,\bp,t) 
{\partial f(\br,\bp,t) \over \partial t} 
=  E_{c}(t) {\partial \tilde N(t) \over \partial t} \; .
\eeq
Inserting this result and Eq.~(\ref{eq41}) 
into Eq.~(\ref{eq40}), the latter is easily seen to yield

\bea
{\partial E_{\rm tot} \over \partial t} & = & E_{c}(t) \left( 
{\partial \tilde N \over \partial t} + 
{\partial N_{c} \over \partial t}
\right) \nonumber \\
& = & 0 \; ,
\eea
due to the conservation of total particle number.
Thus the assumption of adiabaticity is sufficient to ensure that the 
total energy is conserved. However, one can also show the
conservation of energy exactly, without assuming adiabaticity, 
by making use of the dissipative nonlinear Schr\"odinger 
equation in Eq.~(\ref{nlse}). 

\section{THOMAS-FERMI APPROXIMATION}
\label{section5}

The assumption that the condensate grows adiabatically implies that the
dynamics of the condensate itself is being neglected, apart from
its trivial time dependent normalization. In particular, 
we are ignoring the possible excitation of internal
collective oscillations. However, at the temperatures of interest in the
growth process, these excitations are strongly damped and we
expect the condensate to remain in a relatively quiescent
state which is well approximated by the quasi-equilibrium solution of
the GP equation. Indeed, in the experiments there is no evidence of
condensate oscillations, although the thermal cloud has been observed to
oscillate at twice the harmonic oscillator frequency of the trap in some
cases.

For a large number of condensate atoms, a good 
approximation to the equilibrium wave function is provided by the 
Thomas-Fermi approximation which neglects the kinetic energy in the GP 
equation. In this situation, the condensate density is given by
\beq
n_c(\br,t) = {1\over g} \Big [ \mu_c(t) - U_{\rm ext}(\br) - 2g \tilde 
n(\br,t) \Big ]\,.
\eeq
Of course, this expression is only valid if the right-hand side 
is larger than zero; otherwise, $n_c(\br,t)=0$.
The last term on the right-hand side reflects the mean-field interaction
of the condensate with the thermal cloud. Since the latter has a small
density relative to the condensate, its effect on the spatial
distribution of the condensate is small ($2 g \tilde n \ll \mu_c$) and
we therefore neglect it when determining the condensate density. By the
same token, we shall neglect the mean-field interaction of the 
noncondensate with itself. Strictly speaking, these approximations lead
to a violation of total energy conservation, but the error will be very
small since the bulk of the mean-field energy, which resides within 
the condensate itself, is still taken into account. In principle,
these contributions can be included in our treatment as shown explicitly
in the Appendix. However, because these corrections are
small, we have decided to neglect them in our numerical calculations.

It should be noted that the Thomas-Fermi approximation is to 
some extent dictated by our
semi-classical treatment of the noncondensate atoms, since it avoids a
potential problem associated with the placement of the condensate 
chemical potential $\mu_c$ relative to the minimum energy 
 available to the thermal atoms, i.e.,  
$U_{\rm min}=\min[U_{\rm ext}+2g(\tilde n+n_c)]$. For small
condensate densities, it is possible that the GP eigenvalue $\mu_c$ lies
{\it above} this minimum value which is clearly impossible if a full
quantum treatment of the excited states is retained. In  the
Thomas-Fermi approximation there is no such problem since the chemical
potential is exactly equal to $U_{\rm min}$.

Given these approximations, 
the time-dependent condensate density profile becomes

\bea
\label{TFdensity}
n_{c}({\bf r},t) =
{1 \over g} \left[
\mu_{c}(t) - U_{\rm ext}({\bf r}) \right]\,,
\eea
where the external potential is taken to be
a general anisotropic harmonic confining potential,
$U_{\rm ext}({\bf r}) = \sum_{i} m \omega_{i}^{2} r_{i}^{2}/2$.
This expression for the density is again only meaningful when
$U_{\rm ext}({\bf r}) \le \mu_{c}(t)$.
The chemical potential of the condensate is given by

\bea
\mu_{c}(t)={\hbar \bar \omega \over 2} 
\left[ 15 N_{0}(t) {a \over l} \right]^{2/5} \; ,
\eea
where $\bar \omega = (\omega_{1} \omega_{2} \omega_{3})^{1/3}$
and $l=\sqrt{\hbar / m \bar \omega}$.
The potential experienced by the thermal atoms is then
\beq
U(\br,t) 
= \cases{2\mu_c(t) - U_{\rm ext}(\br), &if $n_c \ne 0$\cr
           U_{\rm ext}(\br), &if $n_c = 0$\,.\cr}
\eeq
The minimum value of this potential is $\mu_c(t)$ and occurs on the
boundary of the condensate.

Three additional important quantities can also be calculated
analytically. The first two
are the density of states, and the weighted density of states,
i.e., $\rho(\bar \epsilon)$ and $\rho_{\rm w}(\bar \epsilon)$ 
respectively. For the former we get

\begin{eqnarray}
\rho(\bar \epsilon) & = &
{m^{3/2} \over \sqrt{2} \pi^{2} \hbar^{3}}
\int_{\bar U < \bar \epsilon} d{\bf r} \sqrt{\bar \epsilon
- \bar U({\bf r},t)} \nonumber \\
& = & {2 \over \pi \hbar \bar \omega}
\left[
\int_{\bar U < \bar \epsilon}
dy y^{2} \theta \left({2 \mu_{c} 
\over \hbar \bar \omega} - y^{2} \right)
 \sqrt{{2 (\bar \epsilon  - 
\mu_{c}) \over \hbar \bar \omega} + y^{2}} \right . \nonumber\\
&&\hskip .5truein \left.+ 
\int_{\bar U < \bar \epsilon}
dy y^{2} \theta \left(y^{2} - 
{2 \mu_{c} \over \hbar \bar \omega} \right) 
\sqrt{{2 (\bar \epsilon + \mu_{c}) \over \hbar \bar \omega} - y^{2}}
\,\right] \nonumber \\
& \equiv &
{2 \over \pi\hbar \bar \omega}
\left[
I_{-}(\bar \epsilon) + I_{+}(\bar \epsilon)
\right]
\end{eqnarray}
The integrals $I_{-}(\bar \epsilon)$ and $I_{+}(\bar \epsilon)$
are standard, and are given by

\begin{eqnarray}
I_{-}(\bar \epsilon) & = & \left. 
{u_{-}^{3} x\over 4}
- {a_{-} u_{-} x\over 8} - {a_{-}^{2} \over 8} \log(x + u_{-}) 
\right|_{x=\sqrt{\max \{ 0, 
a_{-} \} }}
^{x=\sqrt{2 \mu_{c} / \hbar \bar \omega}}
\nonumber \\
I_{+}(\bar \epsilon) & = & \left. 
- {u_{+}^{3} x\over 4}
+ {a_{+} u_{+} x\over 8} 
+ {a_{+}^{2} \over 8} \arcsin
\left( {x \over \sqrt{a_{+}}} \right)
\right|_{x=\sqrt{2 \mu_{c} / \hbar \bar \omega}}
^{x=\sqrt{a_{+}}}
\end{eqnarray} 
where we have defined 
$a_\pm=2 (\bar \epsilon \pm \mu_{c})/ \hbar \bar \omega$,
and $u_{\pm}=\sqrt{ a_{\pm} \mp x^{2} }$.

To obtain  an analytic expression for 
the weighted density of states 
$\rho_{\rm w}(\bar \epsilon)$ we note that

\begin{eqnarray}
{\partial \bar U({\bf r},t) \over \partial t} = 
2 {\partial \mu_{c}(t) \over \partial t} \theta[\mu_{c}(t) - 
U_{\rm ext}({\bf r)}]
- {\partial \mu_{c}(t) \over \partial t}\,.
\end{eqnarray}
We therefore find that the weighted density of states is given by

\begin{eqnarray}
\label{weighted}
\rho_{\rm w}(\bar \epsilon) & = & {\partial \mu_{c} \over 
\partial t} \left[ - \rho(\bar \epsilon)
+ {4 \over \pi \hbar \bar \omega} \int_{\bar U < \bar \epsilon}
dy y^{2} \theta \left({2 \mu_{c} \over \hbar \bar \omega} 
- y^{2} \right) \sqrt{{2 (\bar \epsilon - \mu_{c})
\over \hbar \bar \omega} + y^{2}}
\,\right] \nonumber \\
& = & {\partial \mu_{c} \over \partial t} \left[ 
{4 \over \pi \hbar \bar \omega} I_{-}(\bar \epsilon) -
\rho(\bar \epsilon) \right]
\end{eqnarray} 

The third important quantity that can be calculated
analytically in the Thomas-Fermi approximation
arises in the ergodic projection of $C_{12}$. 
With the variable change in Eq.(\ref{variable change}), and
noting that $U_{\rm min}(t) = \mu_{c}(t)$ in the Thomas-Fermi
approximation, Eq.~(\ref{I12}) can be written as

\begin{eqnarray}
\label{I12shifted}
I_{12}(\bar \epsilon_1) & = &
{m^3 g^2 \over 2\pi^3\hbar^7} 
\int d \bar \epsilon_2 \int d \bar \epsilon_3
\int  d \bar \epsilon_4\, 
\left[ \delta( \bar \epsilon_1- \bar \epsilon_2)-
\delta( \bar \epsilon_1- \bar \epsilon_3)-
\delta( \bar \epsilon_1- \bar \epsilon_4) \right]
\nonumber \\ &&\times
\left [ (1+g_2)g_3 g_4 - g_2 (1+g_3)(1+g_4)
\right ]\nonumber \\ &&\times 
\int_{ \bar U\le  \bar \epsilon_{\rm min}} 
d\br\, n_c(\brt) S(p_2,p_3,p_4)\,
\delta(\bar \epsilon_2- \bar \epsilon_3- \bar \epsilon_4)
\end{eqnarray}
where $p_{i}=\sqrt{2 m (\bar \epsilon_{i} - \bar U)}$.
It is apparent that the integrand is symmetric
in the variables $\bar \epsilon_{3}$ and $\bar \epsilon_{4}$.
We can therefore assume, without loss of generality that
$\bar \epsilon_{2} \ge \bar \epsilon_{3} \ge \bar \epsilon_{4}$
which also implies $p_{2} \ge p_{3} \ge p_{4}$. In this
situation, $S(p_{2}, p_{3}, p_{4})$ in Eq.(\ref{Sp2p3p4})
reduces to

\begin{eqnarray}
S(p_{2},p_{3},p_{4}) = {1 \over 2} \left[
1 - {\rm sgn}(p_{2} - p_{3} - p_{4})
\right] \; ,
\end{eqnarray}
which is nonzero and equal to $1$, if 

\begin{eqnarray}
\label{inequality}
p_{2} < p_{3} + p_{4}\,.
\end{eqnarray}
This restricts the spatial integration
in Eq.~(\ref{I12shifted}) 
to the domain specified by this inequality.
Inserting the definitions of $p_{i}$ and
using the conservation of energy condition, $\bar \epsilon_{2} =
\bar \epsilon_{3} + \bar \epsilon_{4}$,
Eq.~(\ref{inequality}) is equivalent to

\begin{eqnarray}
F(\bar U) & \equiv & 
\bar U^{2} - {4 \over 3} (\bar \epsilon_{3} +
\bar \epsilon_{4}) \bar U + 
{4 \over 3} \bar \epsilon_{3} \bar \epsilon_{4} > 0\,.
\end{eqnarray}
The roots of $F(\bar U)=0$ are given by

\begin{eqnarray}
\bar U_{\pm} = {2 \over 3} \left[
(\bar \epsilon_{3} + \bar \epsilon_{4})
\pm \sqrt{ 
\bar \epsilon_{3}^{2} - \bar \epsilon_{3} \bar \epsilon_{4} +
\bar \epsilon_{4}^{2}} \right]\,,
\end{eqnarray}
in terms of which $F(\bar U) = (\bar U - \bar U_{-})(\bar U +
\bar U_{+})$. The requirement $F(\bar U) > 0$ is therefore satisfied
for $ \bar U < \bar U_{-}$ and $\bar U > \bar U_{+}$. The latter condition, 
however, is inconsistent with the constraint $\bar U < \bar
\epsilon_{4}$ for the integral in Eq.~(\ref{I12shifted}). 
Because $\bar U_{-}$ does satisfy
$\bar U_{-} \le \bar \epsilon_{4}$, the net effect
of the factor $S(p_{2},p_{3},p_{4})$ in Eq.~(\ref{I12shifted})
is to restrict the spatial integration domain to 
the domain defined by $\bar U \le \bar U_{-}$, i.e.,   

\begin{eqnarray}
\label{spatial integral}
\int_{ \bar U \le  \bar \epsilon_{\rm min}}
d\br\, n_c(\brt) S(p_2,p_3,p_4)\,
\delta(\bar \epsilon_2- \bar \epsilon_3- \bar
\epsilon_4) = \int_{ \bar U \le \bar U_{-}}
d\br\, n_c(\brt) \delta(\bar \epsilon_2 - 
\bar \epsilon_3 - \bar \epsilon_4)\,.
\end{eqnarray}
The remaining spatial integral in 
Eq.~(\ref{spatial integral}) can be carried 
out analytically for the Thomas-Fermi density
profile. If $\bar U_{-} \ge \mu_{c}$, we have simply

\begin{eqnarray}
\label{Sfactor1}
\int_{ \bar U \le \bar U_{-}} d\br\, n_c(\brt) & = & N_{c}(t)\,.
\end{eqnarray}
On the other hand, for $0 \le \bar U_{-} \le \mu_{c}$, we have

\begin{eqnarray}
\label{Sfactor2}
\int_{ \bar U \le \bar U_{-}} d\br\, n_c(\brt) & = & 
N_{c}(t) \left\{
{5 \over 2} \left[ 1 - \left(1 - {\bar U_{-} 
\over \mu_{c}} \right)^{3/2} \right]
- {3 \over 2} \left[ 1 - \left(1 - {\bar U_{-} \over \mu_{c}} \right)^{5/2} 
\right] \right\} \; .
\end{eqnarray}
Physically, Eqs.~(\ref{Sfactor1}) and (\ref{Sfactor2}) 
are a consequence of the
kinematical constraints for scattering into the condensate
that appear in the original form of the collision integral
in Eq.~(\ref{I12}).

Finally, we indicate some implications of the assumption of adiabatic
growth in the context of the Thomas-Fermi approximation. We take as an
approximate solution to the dissipative nonlinear Schr\"odinger 
equation in Eq.~(\ref{nlse}) a condensate wavefunction of the form

\begin{eqnarray}
\label{adiabatic}
\Psi(\br,t) = \sqrt{n_{c}({\bf r},t)} e^{i \theta(\br, t)} \; ,
\end{eqnarray}
where $n_{c}({\bf r},t)$ is  the Thomas-Fermi density profile
in Eq.~(\ref{TFdensity}). Inserting this wavefunction
into Eq.~(\ref{nlse}), neglecting the kinetic energy term as in the
Thomas-Fermi approximation, and separating real and imaginary parts
of the resulting equation, we obtain the relations

\begin{eqnarray}
\label{theta}
\theta(\br, t) &=& - {1 \over \hbar} \int_{0}^{t} dt' \,\mu_{c}(t') \; ,
\end{eqnarray}
and

\begin{eqnarray}
\label{R}
R(\br, t) & = & - {\hbar \dot \mu_{c}(t) \over 2 g n_c(\br ,t)} \; .
\end{eqnarray}
The fact that the phase is spatially independent implies that the
superfluid velocity $\bv_{c}$ is zero as we have assumed in 
Sec.~\ref{section3}. According to Eq.~(\ref{eq36}), Eq.~(\ref{R}) 
implies

\beq
{\partial N_c \over \partial t}={1\over g}\int d\br\, \dot \mu_c(t)\,,
\eeq
where the integral is restricted to the region occupied by the
condensate, i.e., $n_{c}(\br,t) \ne 0$. This is the same expression
obtained by taking the time derivative of the integral of
Eq.~(\ref{TFdensity}) over all space. We therefore see that
the wave function in Eq.~(\ref{adiabatic}) is an
internally consistent solution of the dissipative nonlinear
Schr\"odinger equation.

\section{RESULTS}
\label{section6}
In this Section we present the results of our calculations, which were
performed for the situation corresponding to the MIT 
experiments~\cite{stamper98}. These
used $^{23}$Na atoms confined in an axially symmetric trap with harmonic
frequencies of $18.0$ Hz and $82.3$ Hz along and perpendicular to the 
symmetry axis, respectively. These values give an averaged frequency of 
$\bar \omega/2 \pi = 49.6$ Hz, 
which implies that $\hbar \bar \omega / k_{B}$ is equal to $2.4$ nK.
The $s$-wave scattering length $a$ is $2.75$ nm.

To begin, we
provide a few of the numerical details. We used a discretized energy 
mesh consisting of equally spaced points in the range $0 \le \bar 
\epsilon \le \bar \epsilon_{\rm max}$. The
value of the temperature used in the simulations is typically of the
order of $1 \; \mu$K, which requires a maximum energy range of about $\bar
\epsilon_{\rm max} \simeq 2500-3000 \; \hbar\bar \omega$ in order to 
ensure that $\rho(\bar \epsilon) g(\bar \epsilon)$
is sufficiently small at the end of the range. In evaluating the
collision integrals $I_{22}$ in Eq.~(26) and $I_{12}$ in Eq.~(30), the
delta functions were used to perform some of the integrations
analytically. The remaining integrals were then 
evaluated numerically using a simple trapezoidal integration scheme.
The main advantage of this scheme in the case of the $I_{22}$ collision
integral is that the conservation of both particle number and
energy is numerically exact, which in general 
is not the case for higher order integration schemes such as Simpson's 
rule. This conserving property is especially important in simulations
of the condensate growth since a loss of either particles or energy 
due to numerical inaccuracy
would lead to systematic errors in the final equilibrium values for
various physical quantities. The situation for the $I_{12}$ collision
integral is somewhat different since neither of the integrals in
Eq.~(35) or (36) is zero. Thus the numerical results will depend on 
the choice of the energy mesh size, and one must check that the
results obtained for a given simulation are insensitive to variations 
in this parameter. The checks we have performed indicate that errors 
in the final 
results coming from this numerical source are no larger than a few
percent. Errors of this magnitude will not influence the
general conclusions that we make. As a final point, we used the Euler
method to propagate Eq.~(32) in time. The time step was chosen to be 
sufficiently small, typically 0.5 ms,
to ensure the accuracy of the time evolution. 

To start the simulation we begin with an initial nonequilibrium
distribution which is meant to represent the conditions immediately
after the rapid evaporative cooling
quench used in the experiments. Ideally, such a quench starts with an
equilibrium distribution at some temperature $T$ above $T_c$ and 
excises all particles with energy above 
$E_{\rm cut} \equiv k_B T_{\rm cut}$. 
We model this by a truncated Bose distribution at the temperature $T$.
Although we expect this initial distribution to represent the 
experimental situation reasonably
well, there will no doubt be differences from the actual
distributions due to the finite time taken to perform the quench, which
allows some equilibration to occur, and the possible incomplete 
removal of
all particles in the energy range of the sweep. Since the rf field is 
resonant only at certain positions in the trap, atoms of a given energy
must have sufficient time to reach these positions in order to suffer
a spin flip and thus be ejected from the trap. If this is not the case,
the distribution in energy will also have a spatial dependence.
Some indication that such a nonergodic state in fact occurs is 
provided by the observation that the thermal cloud starts to
oscillate after the quench. However, for lack of detailed
information about the experimental initial conditions, we shall assume
an idealized truncated Bose distribution as
our initial condition. 

To complete the specification of the initial
state we must also make a choice for the number of atoms initially in
the condensate. Of course, if this number is zero, $I_{12}$ as given by
Eq.~(27) is zero since we have only included stimulated transitions into
the condensate. In the absence of spontaneous processes there is no 
possibility of condensate growth. Under the experimental conditions of
interest, however, the lowest quantum state initially already has a
rather large thermal occupation and stimulated processes
will dominate. We therefore choose the initial condensate number to be 
given by  the occupation of the lowest harmonic oscillator state at the
temperature of the truncated Bose distribution. This number is typically
of the order of a few hundred particles. As our numerical results
presented below will show, the growth curves are rather insensitive to 
this starting value as long as it is small compared to the final 
equilibrium number of condensate atoms.

In Fig.~1 we show a sequence of growth curves which illustrate the
dependence on the parameter
$T_{\rm cut}$. In this set of simulations we assume that the
temperature of the equilibrium Bose distribution is equal to 
$T_c = 0.765 \; \mu K$ and its chemical potential $\tilde \mu$ is equal to
zero. Before the cut, the gas contains $\tilde N = 40 \times 10^6$ 
thermal atoms and the number of condensate atoms is 
given by $N_c = [\exp(3 \beta \hbar \bar \omega/2)-1]^{-1} = 214$. 
In a particular simulation, the total number of atoms and
the average energy per atom of course depends on the
depth of the energy cut. The growth curves are characterized by an 
initial stage of slow growth during which the truncated Bose
distribution evolves into a quasi-equilibrium distribution, a 
well-defined onset time $t_{\rm onset}$ where a significant increase in 
the rate of growth occurs, and finally a relaxational stage where
the condensate number approaches a final equilibrium value. As the cut 
is made deeper and deeper, this final number at first increases due to 
the decreasing total energy of the initial distribution,
which results in  a lower final temperature. However, at some point the
final number of condensate atoms reaches a maximum and then decreases 
with further deepening of the cut
due to the reduced total number of atoms in the initial distribution.
To distinguish this behavior the growth curves are shown as solid lines
when the final number is increasing with decreasing $T_{\rm cut}$, and 
conversely, by dashed lines when the final number is decreasing. 

Although all the growth curves in Fig.~\ref{fig1} are qualitatively 
similar, it is clear that there are important differences in detail.
For the curves with an increasing equilibrium number of
condensate particles, i.e., the solid curves, both the onset time 
and subsequent relaxation time are seen to decrease with decreasing
$T_{\rm cut}$. However, for the 
curves with an decreasing equilibrium number of
condensate particles, i.e., the dashed curves, the
dependence of both of these times on further decreases in $T_{\rm cut}$ is
much weaker, and they appear to approach limiting values. In order to
quantify this behavior, it is convenient to fit the relaxational part
of the theoretical growth curves to a simple exponential relaxation
\beq
\label{fit}
N_c^{\rm fit}(t) \simeq N_c^{\rm eq}\left (1-e^{-\gamma (t-t_{\rm onset})}
\right )\,, \nonumber
\eeq
where $N_c^{\rm eq}$, $\gamma$ and $t_{\rm onset}$ 
are fitting parameters. This
functional form is found to provide a very good fit to this part of the
theoretical curves.
Fig.~2 summarizes the results for the onset time, $t_{\rm onset}$, and
exponential relaxation rate, $\gamma$, for the particular
simulations presented in
Fig.~1. The onset time decreases from about 100 ms to 20 ms as
$T_{\rm cut}/T_c$ is reduced from 5 to 0.5. At the same time, the relaxation
rate increases from about 6 s$^{-1}$ to 12 s$^{-1}$. 

We have also looked at the dependence of the growth curves on the other
parameters that appear in the theory. In Fig.~3 we show the growth
curves for a range of initial temperatures. Prior to 
the quench, these initial temperatures are larger than $T_c$, and in
each case the chemical potential
is adjusted to provide again 
a total of $40\times 10^6$ atoms in the thermal
cloud. The energy cut and initial number of condensate atoms 
were taken to be $T_{\rm cut}/T_c = 2.5$ and $N_c(0) = 214$, respectively,
and were the same for all the runs. Not surprisingly, we find that the final
equilibrium condensate number decreases with increasing initial 
temperature as a result of the larger average energy per atom. This 
of course also leads to a higher final equilibrium temperature.
However what is somewhat unexpected is the very rapid increase of the 
onset time as the initial temperature is increased. In Fig.~4(a) we
show that a $30\%$ 
variation in $T/T_c$ gives rise to more than a ten-fold variation in
$t_{\rm onset}$, and that these values are typically much larger than those
found using an initial temperature of $T = T_c$. In addition, fig.~4(b) 
shows that the relaxation rate tends to decrease with 
increasing $T/T_c$ and is comparable to the values given in Fig~2. 

In Fig.~5 we show the variation of the growth curves with the initial
number of condensate atoms. In this case, the initial nonequilibrium
distribution is held fixed, corresponding to a Bose distribution with
$\tilde N = 40\times 10^6$, $T=T_c$ and $T_{\rm cut}/T_c = 2.5$. The growth
curve is rather insensitive to the condensate number in the range $10^2
< N_c < 10^4$, but then shows a much stronger dependence in the range
$10^4 < N_c < 10^6$. At the higher end of this range, the initial number
is already visible on the graph and by $N_c = 10^6$ there is no longer
a meaningful onset time. This would correspond to a situation in which 
a significant condensate fraction has already formed by the time the
quench is completed. This kind of behavior is indeed also
seen experimentally under certain conditions.

In order to explain some of these results it is necessary to examine the
time evolution of the distribution function $g(\bar \epsilon,t)$. In
Fig.~6 we show $\ln(g)$ vs. $\bar \epsilon$ for various times after the
quench. At early times the distribution function is equilibrated by 
the scattering of thermal atoms into states above the $E_{\rm cut}$ which 
are initially depleted. To conserve energy, the mean energy of the atoms
below $E_{\rm cut}$ must decrease. In fact, the population of the low 
energy states increases significantly before the onset of rapid
condensate growth. 
This is shown in Fig.~7 where $g(\bar \epsilon,t)$ is plotted
as a function of time for some specific energy values. We see that
$g(\bar \epsilon,t)$ at first increases rapidly, reaches a maximum at a
time very close to the onset time and then relaxes towards its final
equilibrium value of $(e^{\beta_{\rm eq} \bar \epsilon} -1)^{-1}$. 
This behaviour is typical of all situations in which the growth of a 
condensate is observed. This strong correlation of the peak position
in Fig.~7 with the onset time suggests that condensate formation is
triggered by an enhanced low-energy population. 
Before the onset time, we find numerically that $g(\bar \epsilon)$ 
behaves approximately as $(\bar \epsilon)^{-1.63}$, which 
is a stronger singularity than that exhibited by an equilibrium Bose 
distribution with zero chemical potential, and agrees within
our numerical accuracy with the
$(\bar \epsilon)^{-5/3}$ dependence predicted by 
Svistunov~\cite{svistunov}. Regardless of the precise exponent, it seems
that a `super-critical' behavior of the distribution function is a 
precursor to condensate formation~\cite{sacket}.

A useful way to characterize
the time evolution of $g(\bar \epsilon,t)$ is to express it locally
as a Bose distribution
\beq
g(\bar \epsilon,t)={1\over \exp(\beta \bar \epsilon - \tilde \mu) -1}\,,
\eeq
where the two parameters $\beta$ and $\tilde \mu$ are defined by 
fitting this expression to the value of the distribution 
function and its energy derivative. Although the parameters are
treated locally as constants in this procedure, they nevertheless
depend parametrically on the energy variable $\bar \epsilon$.
The local temperature and chemical potential parameters defined in this
way are shown in Fig.~8(a) and (b) at time intervals of
0.05 s for a situation in which the quenched thermal cloud
equilibrates to a final temperature above $T_c$. Both parameters are
seen to be strongly energy dependent at early times but evolve towards
energy-independent values by the end of the simulation. The negative
equilibrium value of the chemical potential corresponds to an 
uncondensed thermal cloud at a temperature of about 1.92 $\mu$K.

A situation in which the quench leads to the formation of a condensate 
is illustrated in Figs.~9(a) and (b). The parameters are plotted at
0.25 s intervals during the
relaxational stage of the growth curve beyond the onset time.
At low energies, the local temperature lies above the final equilibrium
value which reflects the higher temperature of the initial Bose
distribution. However at higher energies, the local temperature is 
lower than the final temperature since the gas in this energy range is
effectively colder as a result of the quench. Fig.~9(b)
shows the corresponding variation of the chemical potential. As a result
of the formation of the condensate, the chemical potential at low 
energies is pinned to zero and then increases at higher energies. 
The deviations of both the local temperature and chemical potential from
their final equilibrium values are seen to relax to zero on a time scale
which is comparable with the {\it relaxational} stage of the 
condensate
growth. This relaxation rate can therefore be attributed to the 
relatively slow equilibration of the local temperature and chemical 
potential of the thermal cloud.

We finally turn to a comparison with experiment. This is shown in
Fig.~10 for the particular case in which the starting number 
of noncondensed atoms is $40 \times 10^6$, as in the simulations 
discussed above, but with the initial number of condensate atoms set to
$N_c(0) = 10^4$.
In the particular experimental run starting with this
total number of atoms before the rf quench, the condensate number is
found to relax to a final number of  $1.2 \times 10^6$ atoms.
According to Fig.~1, there are two values of $T_{\rm cut}$ which will
lead to this final number of condensate atoms, $T_{\rm cut}/T_c = 0.6$
and $T_{\rm cut}/T_c = 5.7$. The results for the deeper cut of 
$T_{\rm cut}/T_c = 0.6$ are shown as curve (b) and are seen to be in
very good agreement with the experimental results. However, we cannot
claim good agreement overall since the total number of atoms after the
quench is only $2.5 \times 10^6$ as compared to the 
experimental number of about $16.0 \times 10^6$ atoms. 
For the shallower cut of $T_{\rm cut}/T_c = 5.7$ shown as curve (c), 
the agreement between the theoretical and 
experimental growth curves is clearly worse in that the theoretical
growth rate is too small. Moreover, the total number of atoms
remaining in the trap is $37.5 \times 10^6$ which is too large by
roughly a factor of 2. Alternatively, one can choose a cut which 
reproduces
the final number of atoms in the trap. In our simulations, this requires
a cut of $T_{\rm cut}/T_c = 1.9$. Although the initial growth rate
agrees with experiment in this case, the final equilibrium number of
condensate atoms is $4.5 \times 10^6$, which is too large by almost a
factor of 4. This final number could be improved by elevating the
starting temperature (recall that these simulations used $T = T_c = 
0.765$ $\mu$K), however as Fig.~3 shows, achieving a four-fold reduction
in the equilibrium condensate number would increase the onset time well
beyond the experimental value. It therefore appears that the present
simulations cannot reproduce all aspects of the experiments
simultaneously. 

Fig.~10 also shows a theoretical
growth curve for the same initial conditions as for curve (b), 
but with mean-field
interactions between the condensate and thermal cloud turned off. 
To elaborate, the potential acting on the thermal cloud is simply the
time-independent trapping potential, and the condensate is taken to have
essentially a delta-function spatial distribution at zero energy. In
this case, the integral of $n_c(\bf r,t)$ in Eq.~(\ref{I12}) is
replaced by $N_c(t)$.  It can
be seen that the qualitative behavior is very similar to the fully
interacting simulation, but that the equilibrium number of condensate
atoms is increased considerably, as expected.

Fig.~11 provides a comparison with another set of experimental results.
In this case the initial number of atoms before the quench is not known
and was therefore taken to be $60 \times 10^6$ in order to optimize
agreement with experiment.  Furthermore, the energy cut was chosen as
$T_{\rm cut}/T_c = 2.5$. This leads to
a final number of $7.3 \times 10^6$ condensate atoms in the trap,
which is approximately the same number as found in the experiment,
$N_{c}=7.2 \times 10^6$.
Although this simulation achieves good agreement 
between theory and experiment for the condensate growth curve, there are
too many unknown variables, including the final number of atoms in the
trap, to know whether or not theory is reproducing experiment.
For this reason, the results in Fig.~11 should simply be viewed as a 
possible fit to the experimental data.

\section{DISCUSSION AND OUTLOOK}

\label{section7}
Our main objective has been to obtain a realistic description
of condensate growth
which takes into account the effects of mean-field
interactions. Within the ergodic approximation for the
noncondensed atoms, and the adiabatic approximation 
for the condensate, the kinetic equation we
obtain is given by Eq.~(\ref{eq19}), and we have used this
equation to perform simulations of condensate growth.
In agreement with earlier work\cite{gardiner97,holland97}, 
we find that 
the growth curves have a well-defined onset time, after which an
exponential relaxation towards equilibrium takes place.
Detailed comparison with the results reported in 
Ref.~\onlinecite{stamper98} shows that certain parameters can be tuned 
in order to achieve agreement with the experimental growth curves.
However it seems impossible with the present simulations to reproduce
the overall equilibrium state of the trapped gas. 

If we attribute the existing discrepancies to theory we must at 
some point reexamine
the two major assumptions made in this work, namely the adiabatic growth
of the condensate and the ergodic evolution of the thermal cloud.
The adiabatic assumption neglects the dynamics of the condensate,
specifically the possibility that collective oscillations are excited
during the growth process. Whether or not this has any important
effect on the rate at which atoms are exchanged between the condensate
and thermal cloud in not known and should be investigated. In the same
vein, oscillations of the thermal cloud seen in the experiments clearly
indicate the non-ergodic state of the gas which in principle might be
important in determining the time scale of equilibration.
However, to answer this question requires a solution of the full 
quantum Boltzmann equation which seems out of reach at the moment.
One cannot of course discount the possibility that there are
uncertainties in the experimental results themselves. Further 
experimental
work is needed to confirm the earlier results and to explore in more
detail the dependences on various parameters such as the initial
temperature of the cloud and the depth of the rf cut.

After completion of this work, a preprint 
by Davis, Gardiner, and Ballagh appeared~\cite{davis99}
which is a continuation of a series of papers by Gardiner
{\it et al.}. It also addresses the
issue of mean-field interactions as affecting the density of states, and
improves on the authors' earlier work by giving a more realistic 
description of the rf quench used in the experiments. Thus, although
there are differences in methodology, the physical basis of their work
and the approximations they make are essentially equivalent to ours. 
As confirmation of this equivalence, their calculations of condensate
growth performed for the initial conditions of Figs.~10 and 11 yield
results which are in quantitative agreement with ours.
The situation considered in Fig.~10 is optimal from a theoretical point
of view since the experimental conditions are best known in this case.
Yet both sets of calculations are unable to reproduce the experimental
results in every detail.

One of the differences between their work and ours concerns the way 
that the
condensate is treated. In our formulation, the condensate is isolated 
explicitly as the macroscopically occupied quantum state, and the
remaining excited states making up the thermal cloud are treated
semiclassically. As a result of this formulation, we have two kinds of
collision integrals, one for thermal atoms scattering amongst each other
and a second for collisions of thermal atoms with the condensate. In the
formulation of Davis, Gardiner, and Ballagh on the other hand, all 
states
including the condensate are treated equivalently and thus only a single
collision integral enters. As a result, the effective collision 
cross-section involving the condensate 
does not depend on time as it does in our formulation.
A second apparent difference has to do with the term involving the
weighted density of states $\rho_{\rm w}$ in Eq.~(\ref{eq19}). This term
arises as a consequence of the time dependence of the 
mean-field interaction. Although Davis, 
Gardiner, and Ballagh also 
deal with a time-dependent density of states, the second term
on the right hand side of Eq.~(\ref{eq15}) does not appear explicitly in
their kinetic equation. However, they account for this term by dividing
phase space into energy bins having widths which are a function of time.
A final difference involves the use of the Bogoliubov 
excitation spectrum in the calculation of their density of states,
instead of the Hartree-Fock dispersion used here. We do not expect this
to affect the condensate growth curves significantly. However, if
quasi-particle excitations are invoked, one should in principle 
also use these states
to calculate the collision integrals~\cite{kirkpatrick85}.
It is not known at present what effect this might have on the
collision rates for the low-lying energy levels.

Finally, we note that the ergodic treatment of the Boltzmann equation 
is a powerful, albeit approximate, method which would allow the 
study of nonequilibrium processes in other situations as well.
Some future applications might include
the nonequilibrium dynamics of fermion-fermion and 
boson-fermion mixtures. Thus far, the problem of evaporative 
cooling in these systems has been studied using a simplified
procedure whereby the distribution function is assumed to be given
by a cut-off equilibrium distribution function~\cite{holland99}. 
A cooling trajectory in phase space is then generated by solving for the
temperature, chemical potential and cut-off energy at each successive
time step. The accuracy of this approach could be checked by solving 
for the entire distribution function following the methods used here.
Another interesting application would be to study a 
nonequilibrium steady state situation in which atoms are continuously
fed into the trapping potential while simultaneously being removed by
a rf-cut~\cite{williams99}. This would be relevant to the study
of steady-state atom lasers.

\acknowledgements
E. Z. acknowledges the FOM for its financial support during a sabbatical
visit to the University of Utrecht, as well as support 
from the Natural Sciences and Engineering Research Council of Canada.
We would like to acknowledge useful discussions with A. Griffin and 
J. Williams, and with C. W. Gardiner regarding the work in
Ref.~\onlinecite{davis99}.

\begin{appendix}

\section{Evaluation of the weighted density of states.}
\label{appendixA}

In this Appendix, we summarize the steps needed to include
in our calculations the effect of the mean-field interactions 
arising from the noncondensed cloud itself.
Referring to Eq.~(\ref{eq16}), we see that we 
must evaluate $\partial U({\bf r},t)/\partial t$. 
This quantity is given by

\beq
\label{A1}
{\partial U(\brt) \over \partial t} = 2g \left ( {\partial
\tilde
n(\brt) \over
\partial t} + {\partial n_c(\brt) \over \partial t} \right ) \,.
\eeq
The time derivative of $\tilde n(\brt)$ can be expressed as

\bea
\label{A2}
{\partial \tilde n(\brt) \over \partial t} &=& {\partial \over
\partial t}
\int {d^{\,3}p \over (2\pi\hbar)^3} \int d\epsilon \,
\delta(\epsilon -
E(\br,\bp,t))g(\epsilon,t)\nonumber \\
&=& \int d\epsilon\, \rho(\br,\epsilon,t) \left [ {\partial
U(\brt)
\over \partial t} {\partial g(\epst) \over \partial \epsilon}
+ {\partial g(\epst) \over \partial t} \right ]\nonumber \\
&=& I(\brt){\partial U(\brt) \over \partial t} + \int
d\epsilon\,
\rho(\br,\epst) {\partial g(\epst) \over \partial t} \; ,
\eea
where we have defined

\beq
\label{A3}
I(\brt) \equiv \int d\epsilon\, \rho(\br,\epsilon,t)
{\partial g(\epst) \over \partial \epsilon} \,.
\eeq
Substituting Eq.~(\ref{A1}) into Eq.~(\ref{A2}), 
the latter can be rearranged to
provide an expression for the time rate of change of $\tilde n(\brt)$ 
in terms of the time rate of change of the 
condensate density $n_{c}(\brt)$ and the 
distribution function $g(\epsilon)$. We find

\beq
{\partial \tilde n(\brt) \over \partial t} = {2gI(\brt) \over
1-2gI(\brt)} {\partial n_c(\brt) \over \partial t} +
\int d\epsilon\, {\rho(\br,\epst)
\over 1-2gI(\brt)} {\partial g(\epst) \over \partial t}\,.
\eeq
Inserting this result into Eq.~(\ref{A1}), we have

\beq
{\partial U(\brt) \over \partial t} = -{2g \over 1-2gI(\brt)}
{\partial
n_c(\brt) \over \partial N_c} {\partial \tilde N \over \partial
t}
+ {2g \over 1-2gI(\brt)} \int d\epsilon\,
\rho(\br,\epst) {\partial g(\epst) \over \partial t}\,.
\eeq
We have here made use of the fact that $n_c(\br,t)$ depends on time
parametrically through $N_c(t)$, so that

\beq
{\partial n_c(\brt) \over \partial t} =
{\partial n_c(\brt) \over \partial N_c} {\partial N_c \over
\partial t}
=-{\partial n_c(\brt) \over \partial N_c} {\partial \tilde N
\over
\partial t} \; .
\eeq
Thus, the weigthed density of states becomes

\bea
\rho_{\rm w}(\epst) &=& \int d^{\,3}r \rho(\br,\epsilon,t) {\partial
U(\brt)
\over \partial t} \nonumber \\
&=& -\int d^{\,3}r \rho(\br,\epsilon,t) \left (
{2g \over 1-2gI(\brt)} {\partial n_c(\brt) \over \partial N_c}
\right )
{\partial \tilde N \over \partial t} 
\nonumber \\ && 
+ \int d^{\,3}r \rho(\br,\epsilon,t) \int d\epsilon^{\,\prime}
{2g\rho(\br,\epsilon^{\,\prime},t) \over 1-2gI(\brt)}
{\partial g(\epsilon^{\,\prime},t) \over \partial t}
\nonumber \\
&\equiv& A(\epst) {\partial \tilde N \over \partial t} + \int
d\epsilon^{\,\prime} B(\epsilon,\epsilon^{\,\prime},t)
{\partial g(\epsilon^{\,\prime},t) \over \partial t}\,,
\eea
where

\beq
A(\epst) \equiv - \int d^{\,3}r \rho(\br,\epsilon,t) \left (
{2g \over 1-2gI(\brt)} {\partial n_c(\brt) \over \partial N_c}
\right )\,,
\eeq
and

\beq
B(\epsilon,\epsilon^{\,\prime},t) \equiv 2g\int d^{\,3}r \,
{\rho(\br,\epsilon,t)
\rho(\br,\epsilon^{\,\prime},t) \over 1-2gI(\brt)}\,.
\eeq
We recover the expression for $\rho_{\rm w}(\epst)$ given in
Eq.~(\ref{weighted}) by setting the kernel $B$ equal to zero and
neglecting $I$ in the expression for $A$. It can be seen that including
the mean-field of the noncondensate complicates the calculations
considerably, but all quantities can in principle be calculated
explicitly if these refinements are desired. However, as discussed in
Sec.~\ref{section5}, we do not expect these effects to be
quantitatively important.

\end{appendix}

\begin{figure}[h]
\vspace{0.25 cm}
\caption{\label{fig1}
Growth curves for different initial energy cutoffs.
As discussed in Sec.~\ref{section4}, the initial conditions are 
defined by fixing the temperature $T=T_{c}=0.765\, \mu$K
and the chemical potential $\tilde \mu = 0$ of the distribution 
function before it is truncated. This gives $\tilde N = 40 \times 10^6$
noncondensate atoms. The number of atoms initially in the condensate is
chosen to be $N_c =214$.  The solid curves in 
order of {\it increasing} saturation values correspond to
$T_{\rm cut}/T_c = 5.5$, 5.0, 4.5, 4.0, 3.5, 3.0 and 2.5. The dashed 
curves in order of {\it decreasing} saturation values correspond to 
$T_{\rm cut}/T_c = 2.0$, 1.5, 1.0 and 0.5.
}
\end{figure}

\begin{figure}[h]
\vspace{0.25 cm}
\caption{\label{fig2}
The onset time (a) and relaxation rate (b) for the growth curves in 
Fig.~\ref{fig1}, as determined by using the fitting function in 
Eq.~(\ref{fit}).}
\end{figure}

\begin{figure}[h]
\vspace{0.25 cm}
\caption{\label{fig3}
Growth curves for different initial temperatures $T$. In order of {\it
decreasing} equilibrium number of condensate atoms, 
$T/T_c =$ 1, 1.05, 1.1, 1.15, 1.2, 1.25
and 1.3.  The initial conditions are defined by
fixing the number of noncondensed
particles to $\tilde N = 40 \times 10^{6}$, and the
number of condensed atoms to $N_{c} = 214$, as in Fig.~\ref{fig1}.
The cutoff is now kept fixed at $T_{\rm cut}/T_c=2.5$.
The chemical potential is less than zero, and adjusted to keep the 
number of noncondensed particles fixed.
} 
\end{figure}

\begin{figure}[h]
\vspace{0.25 cm}
\caption{\label{fig4}
The onset time (a) and relaxation rate (b) for the growth curves in 
Fig.~\ref{fig3}, using the same fitting procedure as in 
Fig.~\ref{fig2}.}
\end{figure}

\begin{figure}[h]
\vspace{0.25 cm}
\caption{\label{fig5}
Growth curves for different initial number of condensed
particles. The other parameters defining the initial conditions are the
same as in Fig.~\ref{fig1}, and the cutoff is held fixed at 
$T_{\rm cut}/T_c=2.5$. In order of {\it increasing} saturation values, the
curves correspond to $N_c(0) = 10^2,\, 10^3,\, 10^4,\, 10^5$ and $10^6$.
} 
\end{figure}

\begin{figure}[h]
\vspace{0.25 cm}
\caption{\label{fig6}
Plot of the logarithm of the distribution function 
$g(\epsilon,t)$ for the curve with $T_{\rm cut}/T_{c}=2.5$ from 
Fig.~\ref{fig1}, at time intervals $\Delta t= 0.02$ s, starting from 
$t= 0.02$ s. 
Each curve is shifted up by one unit with respect
to the previous one for clarity. }
\end{figure}

\begin{figure}[h]
\vspace{0.25 cm}
\caption{\label{fig7}
Plot of the product of the density of states and the 
distribution function at energies of
30, 60, 120 and 240 $\hbar \bar \omega$, 
for the curve with $T_{\rm cut}/T_{c}=2.5$ in
Fig.~\ref{fig1}. The peaks occur in the vicinity of the onset time 
$t_{\rm onset}$.}
\end{figure}

\begin{figure}[h]
\vspace{0.25 cm}
\caption{\label{fig8}
Equilibration of the local temperature and chemical potential for a 
situation in which the final
equilibrium temperature of the gas is above the critical temperature. 
Panel (a) gives the local temperature as a function of energy
for a sequence of times during the equilibration process. In
equilibrium, both the temperature and chemical potential are
independent of energy. Panel (b) gives
the corresponding variation of the local chemical potential.
The initial conditions before the distribution is truncated are
defined by a temperature $T=2 \mu K$, and a chemical potential 
$\tilde \mu=-200$. The cutoff is at $T_{\rm cut}/T_c = 2.5$.} 
\end{figure}

\begin{figure}[h]
\vspace{0.25 cm}
\caption{\label{fig9}
As in Fig.~\ref{fig8}, but for a situation in which the final 
equilibrium temperature is below the critical temperature. The
equilibration of the local temperature and chemical potential 
corresponds to the $T_{\rm cut}/T_c = 2.5$ growth curve
in Fig.~\ref{fig1}.
}
\end{figure}

\begin{figure}[h]
\vspace{0.25 cm}
\caption{\label{fig10}
Theoretical growth curves for the initial conditions of Fig.~\ref{fig1},
but with $N_c(0) = 10^4$ and for various energy cuts: (a) 
$T_{\rm cut}/T_c = 1.9$, (b) $T_{\rm cut}/T_c = 0.6$, (c)
$T_{\rm cut}/T_c = 5.7$. The experimental points are taken
from Fig.~\ref{fig4} of Ref.~[10]. The dashed line shows the
theoretical growth curve for the conditions of case (b)
but with mean field interactions turned off.
}
\end{figure}

\begin{figure}[h]
\vspace{0.25 cm}
\caption{\label{fig11}
Theoretical growth curve for initial conditions
given by $\tilde N(0) = 60\times 10^6$, $T=T_c =
0.876 \; \mu$K, $T_{\rm cut}/T_c = 2.5$ and $N_c(0) = 50 \times 10^4$. The 
experimental points are taken from Fig.~\ref{fig3} of Ref.~[10].
}
\end{figure}

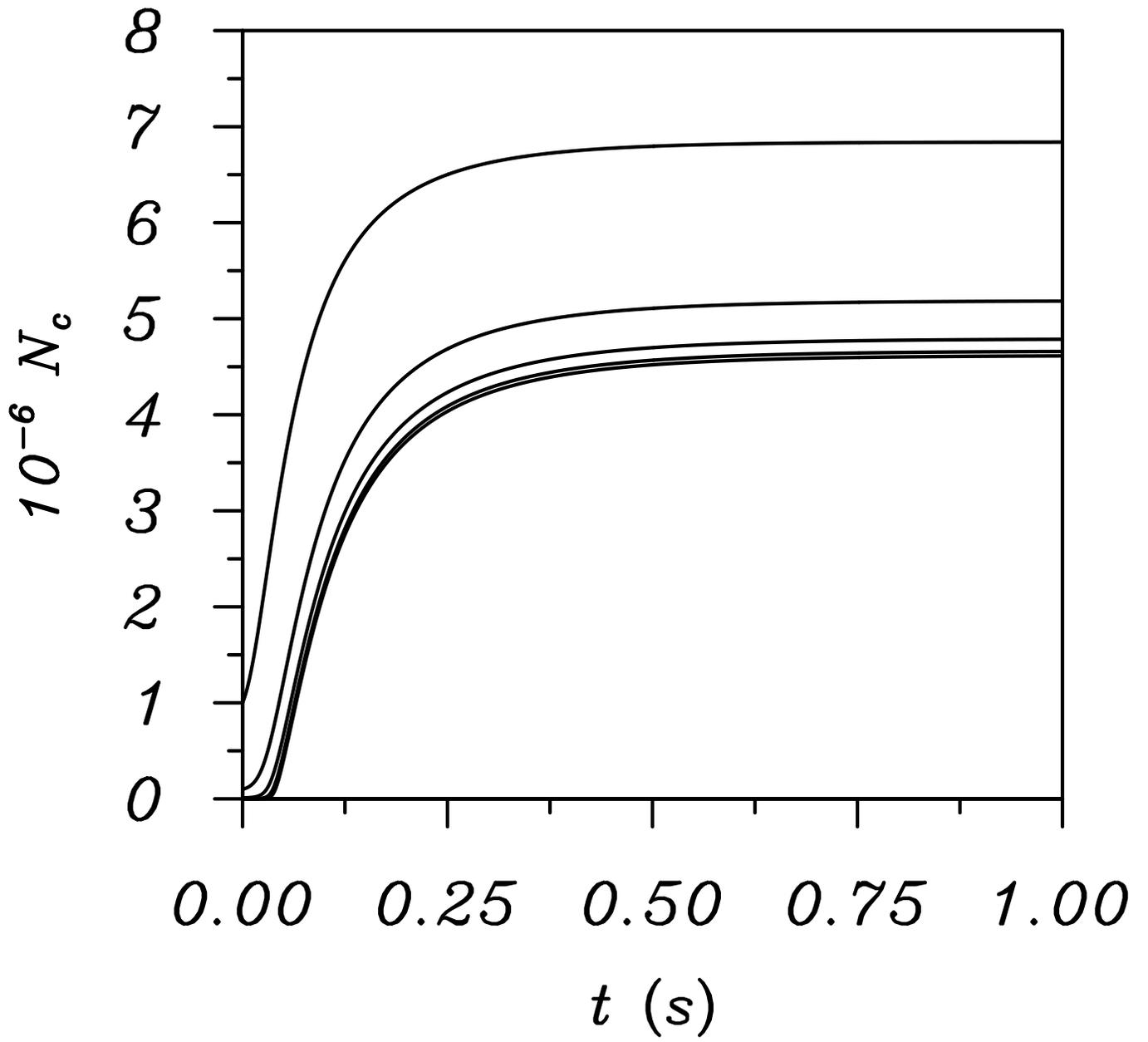
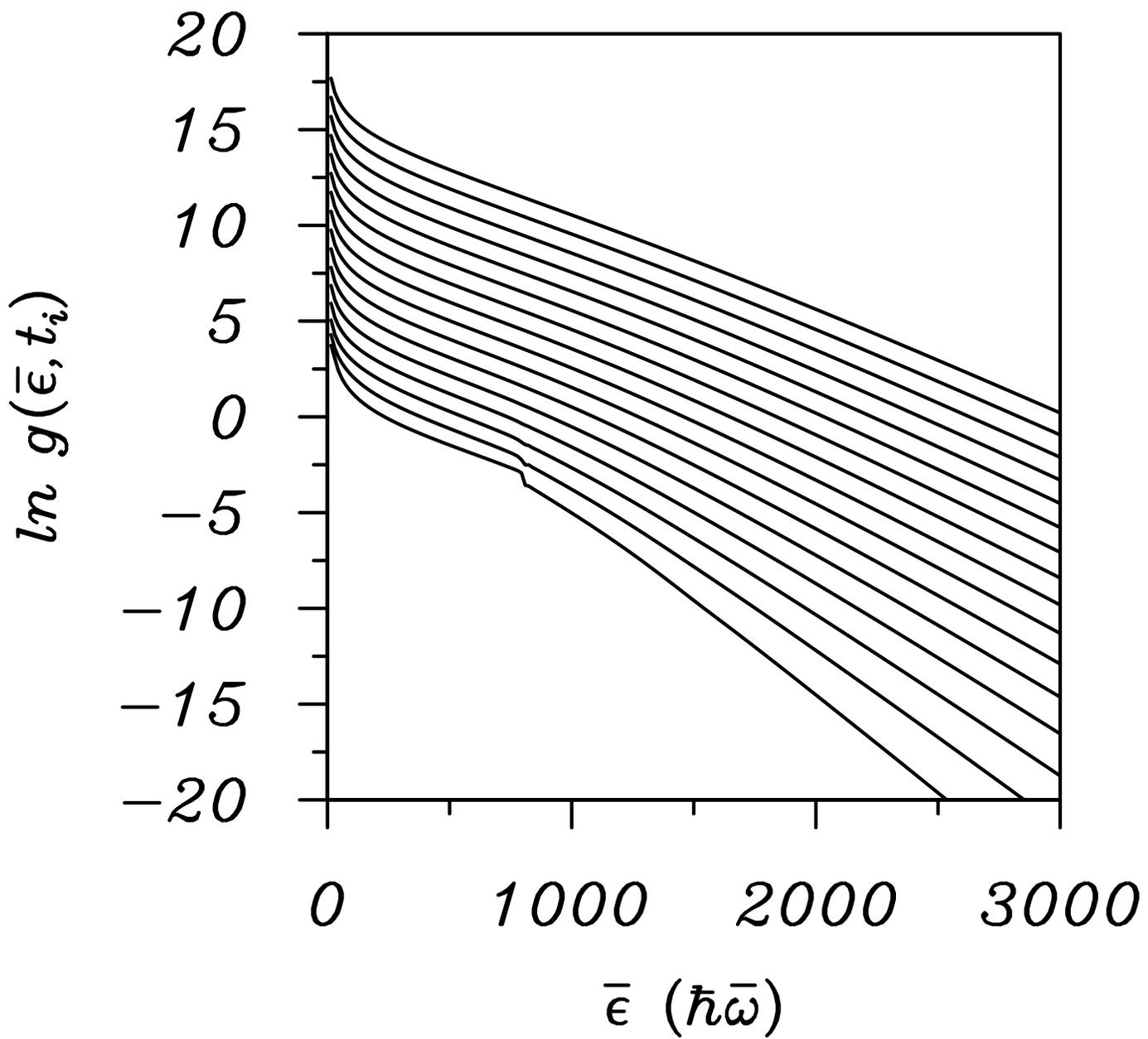
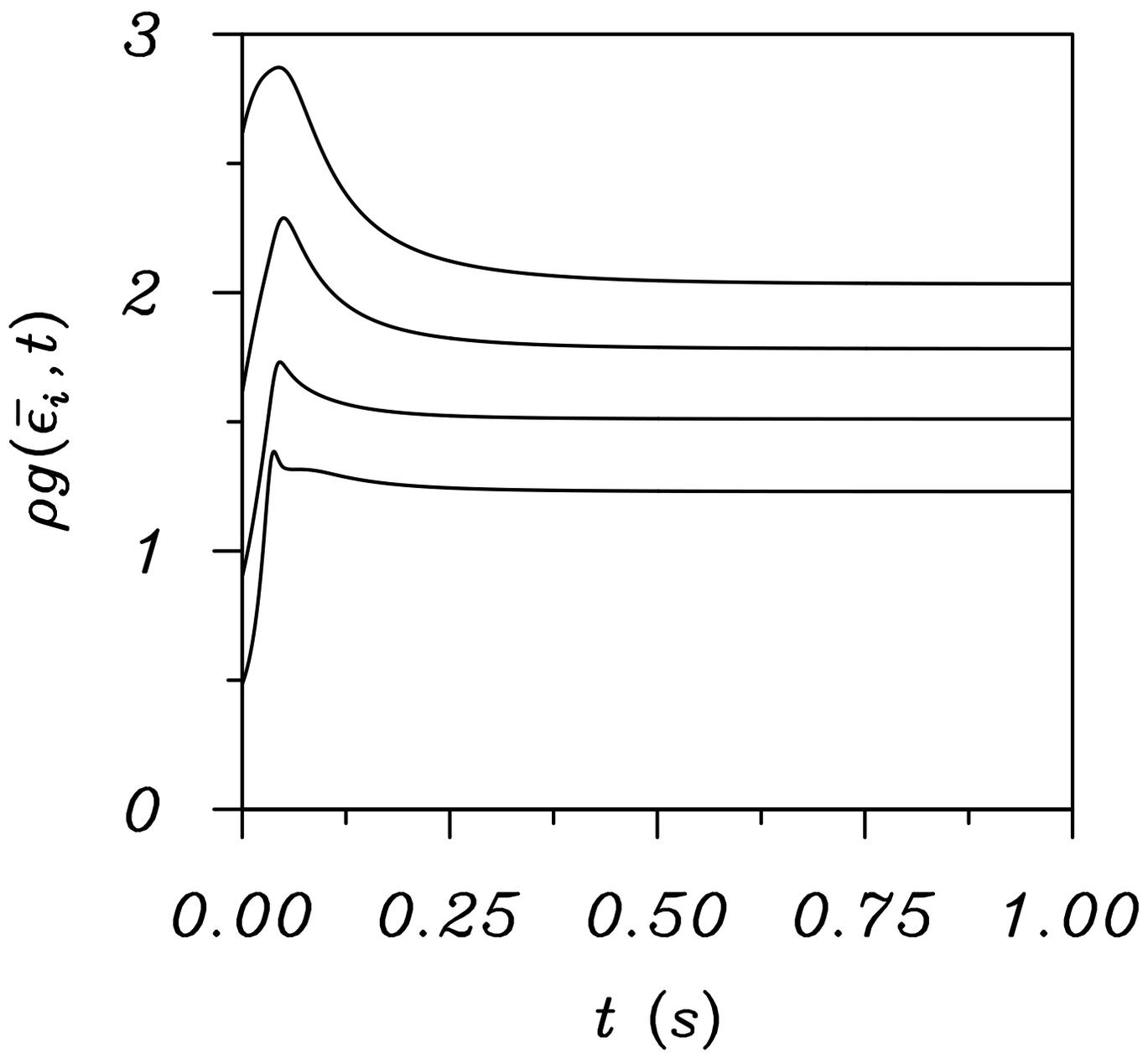
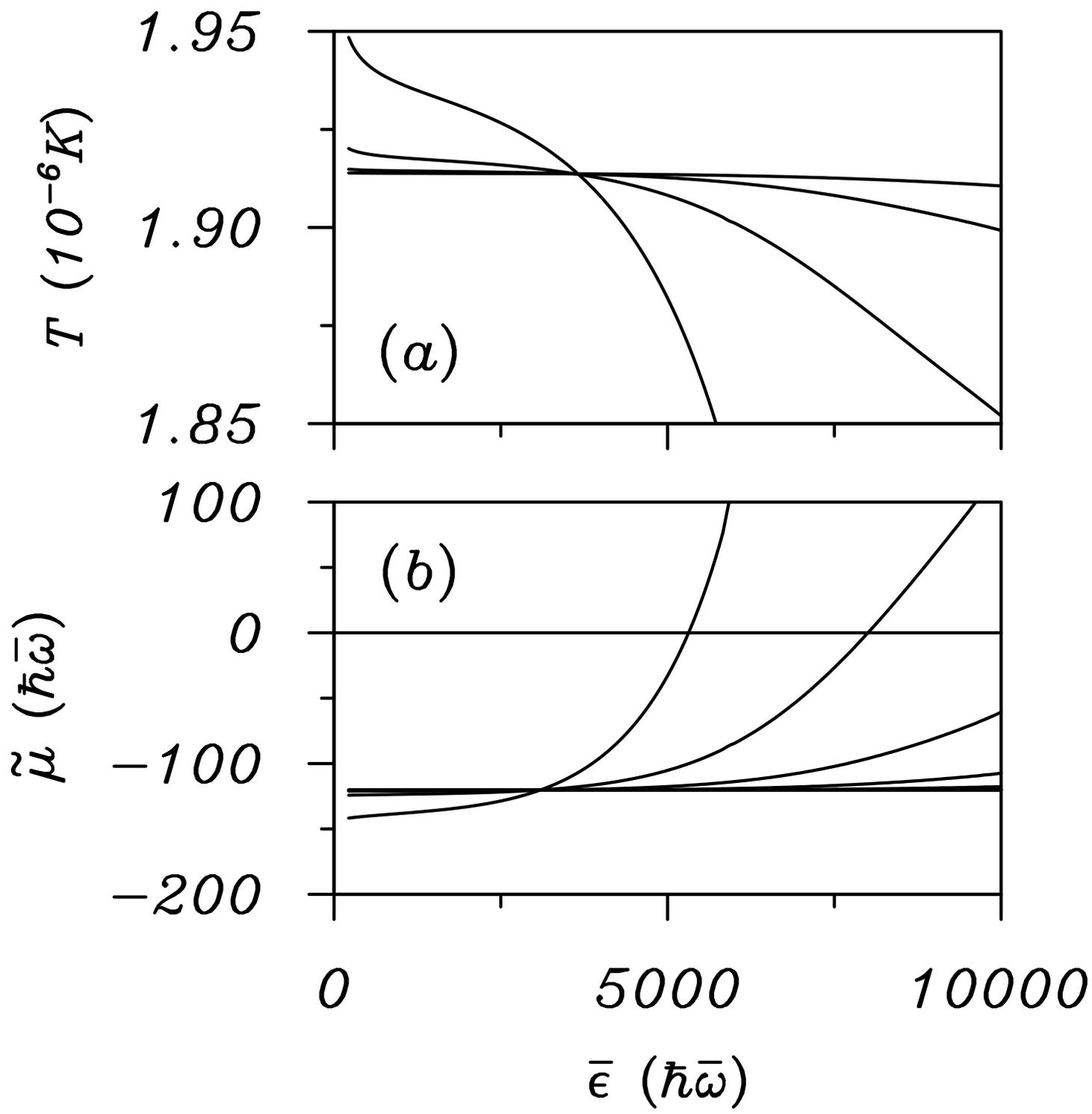
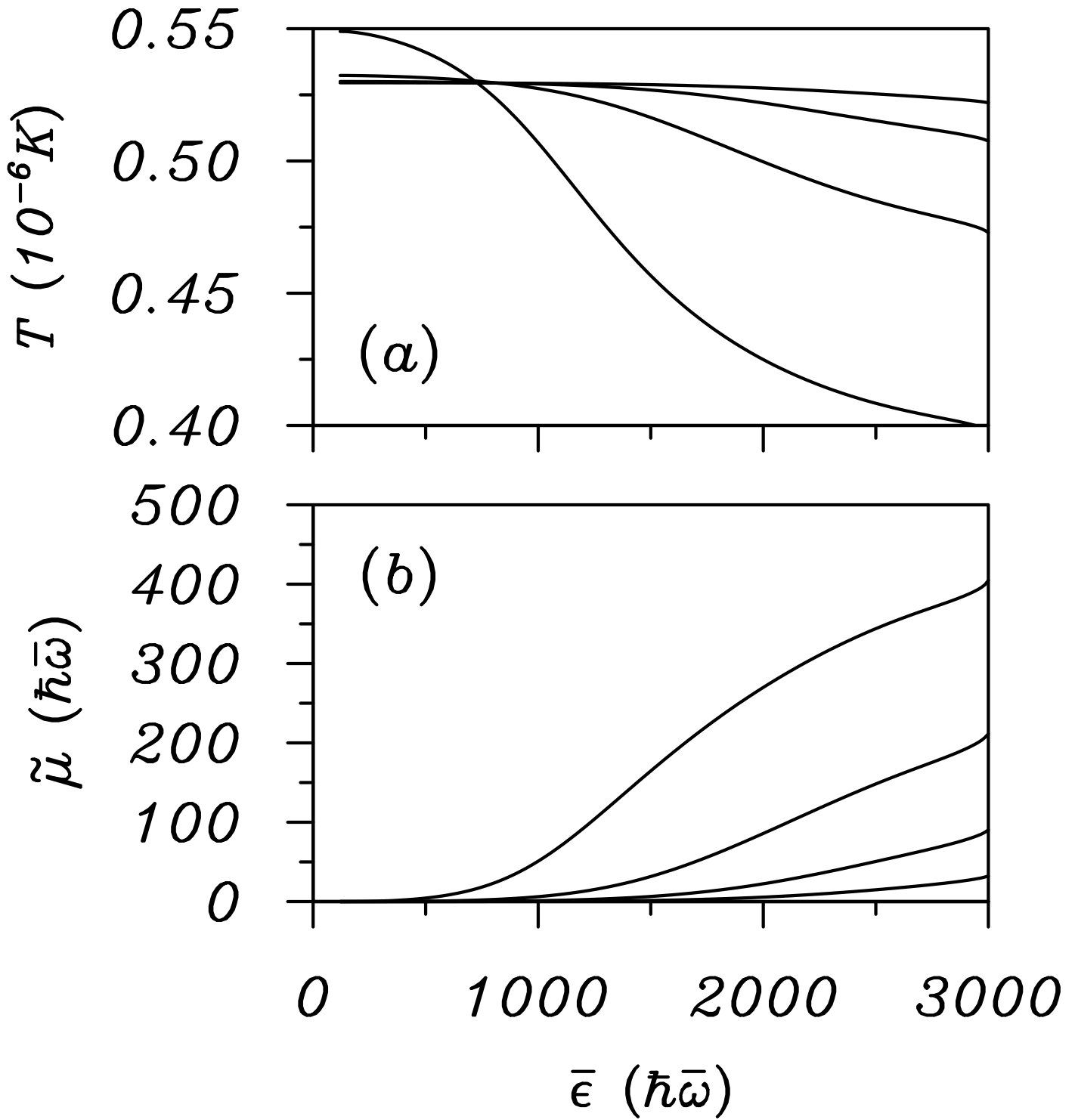
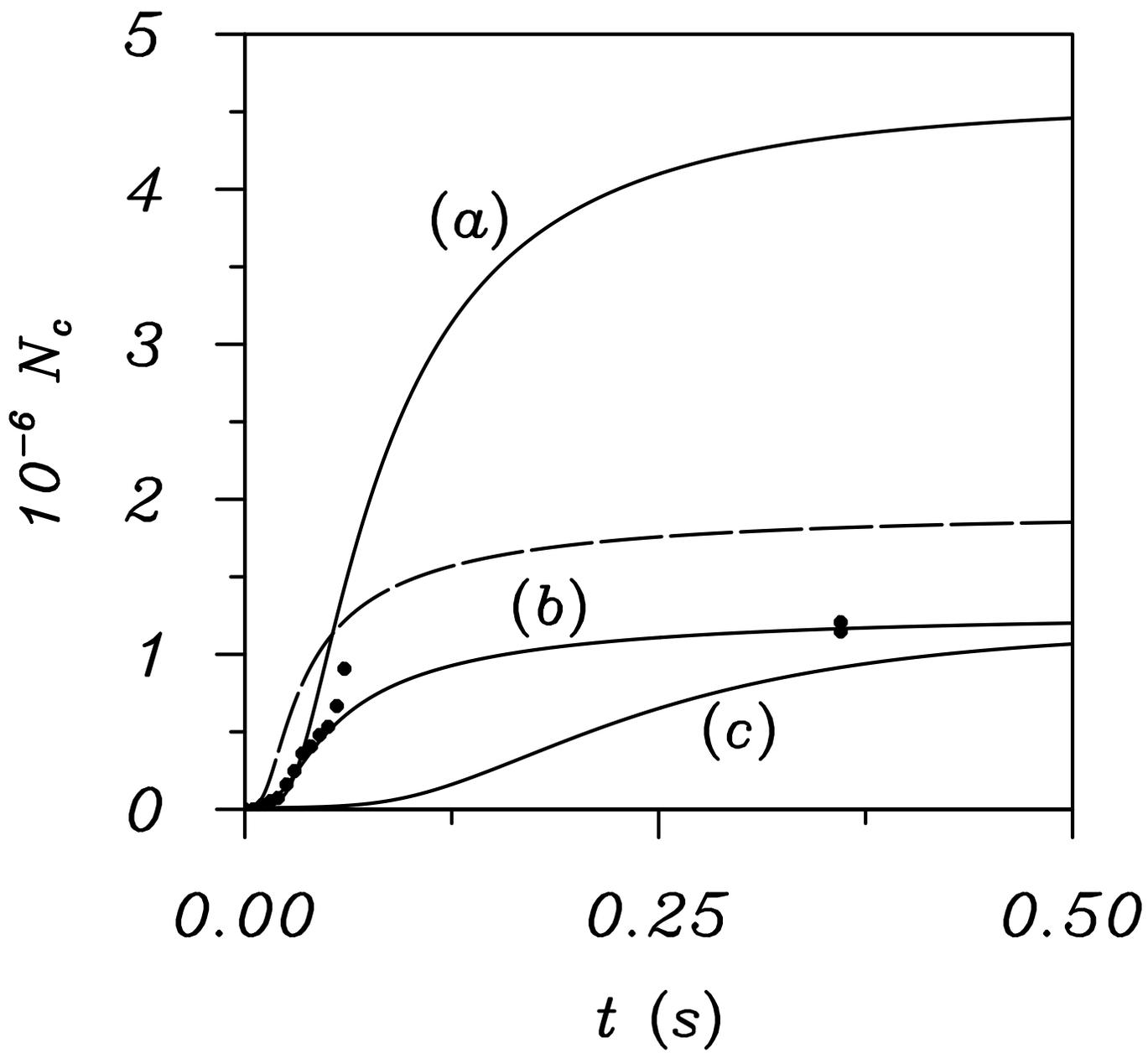
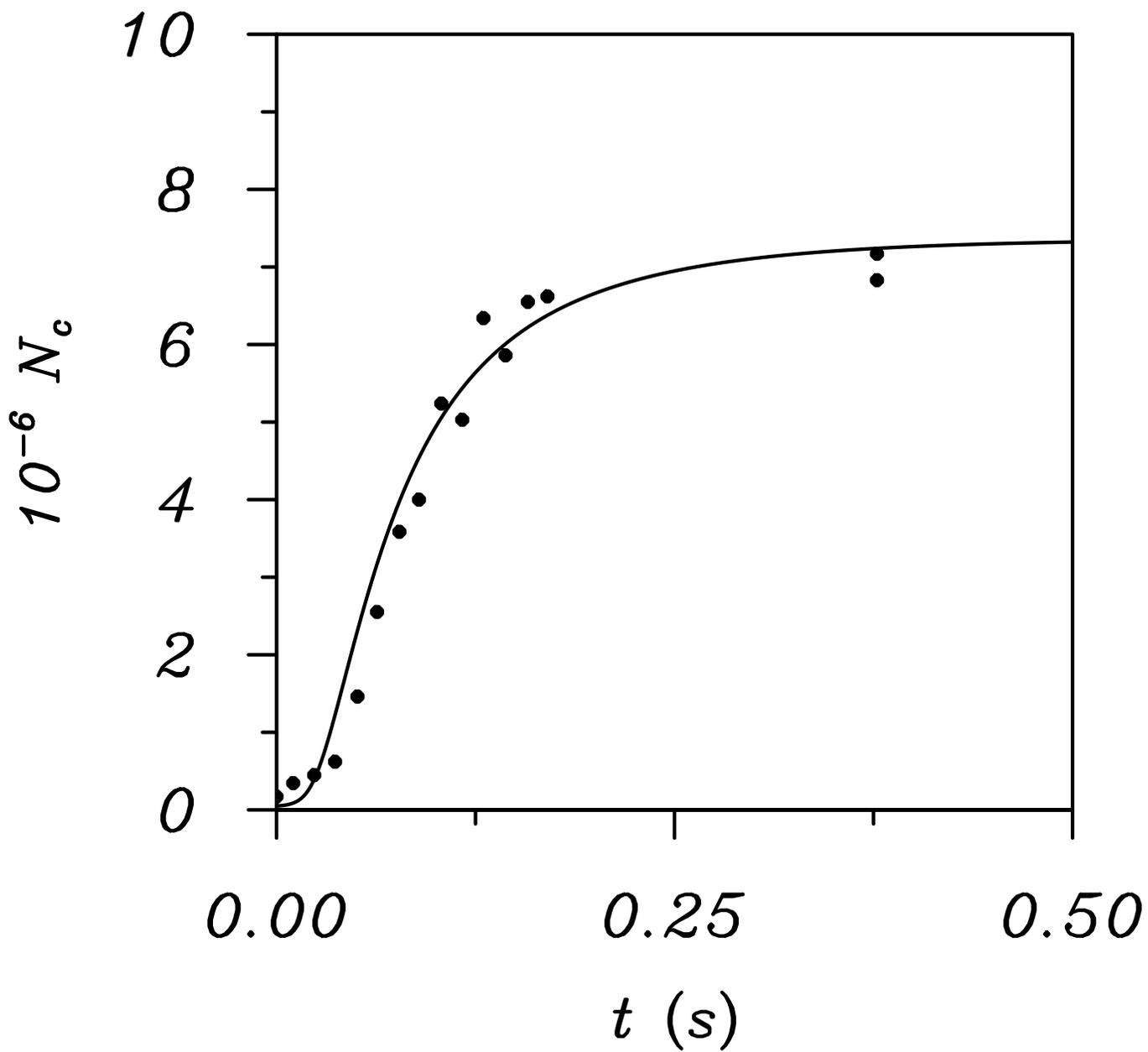
\end{document}